\documentclass[%
reprint,
superscriptaddress,
showpacs,preprintnumbers,
amsmath,amssymb,
prb,
]{revtex4-2}
\usepackage{graphicx}
\usepackage{dcolumn}
\usepackage{xfrac}
\usepackage{bm}
\usepackage{color}
\usepackage[export]{adjustbox}
\usepackage{xr-hyper}
\usepackage{hyperref}
\usepackage{tabularx}

\usepackage{amsmath}
\usepackage{amsfonts}
\usepackage{amssymb}
\usepackage{chemformula}
\usepackage{soul,xcolor}
\usepackage{lipsum}
\setstcolor{red}

\newcommand{\TC}{$T_{\text C}$}

\newcommand{\degree}{^{\circ}}
\newcommand{\angstrom}{\text{\normalfont\AA}}

\begin{document}

\title{A high-entropy form of \ch{$R$Mn6Sn6} with distinct magnetotransport regimes correlated to different magnetic structures}

\author{Kyle~W.~Fruhling$^{\ast\dagger}$}
\affiliation{Department of Physics, Boston College, Chestnut Hill, MA 02467, USA}

\author{Jonathan~Gaudet}
\affiliation{NIST Center for Neutron Research, National Institute of Standards and Technology, Gaithersburg, MD 20899, USA}
\affiliation{Department of Materials Science and Eng., University of Maryland, College Park, MD 20742, USA}

\author{William~D.~Ratcliff}
\affiliation{NIST Center for Neutron Research, National Institute of Standards and Technology, Gaithersburg, MD 20899, USA}
\affiliation{Department of Physics and Department of Materials Science and Engineering, University of Maryland, College Park, MD 20742, USA}

\author{Jonathan~S.~White}
\affiliation{PSI Center for Neutron and Muon Sciences, Forschungsstrasse 111, 5232 Villigen, PSI, Switzerland}

\author{Siddharth~Nandanwar}
\affiliation{Department of Physics, Boston College, Chestnut Hill, MA 02467, USA}

\author{Noah~J.~Fau}
\affiliation{Department of Chemistry and Biochemistry, Baylor University, Waco, TX 76798, USA}

\author{Gregory~T.~McCandless}
\affiliation{Department of Chemistry and Biochemistry, Baylor University, Waco, TX 76798, USA}

\author{Enrique~O.~Gonz\'{a}lez-Delgado}
\affiliation{Department of Physics, Boston College, Chestnut Hill, MA 02467, USA}

\author{Julia~Y.~Chan}
\affiliation{Department of Chemistry and Biochemistry, Baylor University, Waco, TX 76798, USA}

\author{Oksana~Zaharko}
\affiliation{PSI Center for Neutron and Muon Sciences, Forschungsstrasse 111, 5232 Villigen, PSI, Switzerland}

\author{Michael~A.~Susner}
\affiliation{Foundational Technologies Directorate, Air Force Research Laboratory, Wright-Patterson Air Force Base, Ohio 45433, USA}

\author{Michael~J.~Newburger}
\affiliation{Foundational Technologies Directorate, Air Force Research Laboratory, Wright-Patterson Air Force Base, Ohio 45433, USA}

\author{Fazel~Tafti}
\email{fazel.tafti@bc.edu}
\affiliation{Department of Physics, Boston College, Chestnut Hill, MA 02467, USA}


\begin{abstract}

The kagome \ch{$R$Mn6Sn6} material family has attracted significant attention as high-temperature metallic magnets with a host of different magnetic orderings and anisotropy. 
Theoretical studies point to the rare-earth ($R$) as the determining factor for both the direction of magnetic anisotropy and the type of magnetic ordering in a given compound.
This motivates studying high-entropy forms of \ch{$R$Mn6Sn6} to examine how the interplay of several rare-earth elements leads to different magnetic states in a single crystal. 
Here, we present a rare-earth mix of Tb, Dy, Ho, Er, Tm, and Lu that produces phase transitions from a paramagnet to an easy-plane ferrimagnet (FiM) below \TC~= 380 K, then to a FiM easy-axis state at $T_{\text{SR1}}$ = 207 K, to a canted FiM ground state below $T_{\text{SR2}}$ = 79 K.
This behavior is consistent with previously reported high-entropy \ch{$R$Mn6Sn6} compounds; however, uniquely, the rare-earth mix studied here exhibits a broad transition from easy-plane to easy-axis anisotropy from 270 K to 170 K, and reveals a nonmonotonic magnetoresistance.
Using neutron scattering data, we found that both observations correlate with an incommensurate modulated contribution to the spin state due to competing rare-earth interactions.
This magnetoresistive behavior and the correlated spin structures underscore the potential for rare-earth engineering of magnetism. 

\end{abstract}

\maketitle


\section{\label{sec:introduction}Introduction}

Kagome lattice materials have received significant attention due to the potential of their electronic states to host flat bands, Dirac crossings, and van Hove singularities~\cite{yin_topological_2022,wang_quantum_2023}. 
Incorporating magnetic elements into a kagome lattice enables further tuning of their electronic and magnetic states~\cite{liu_magnetic_2026}.

\begin{figure*}
\centering
  \includegraphics[width=\textwidth]{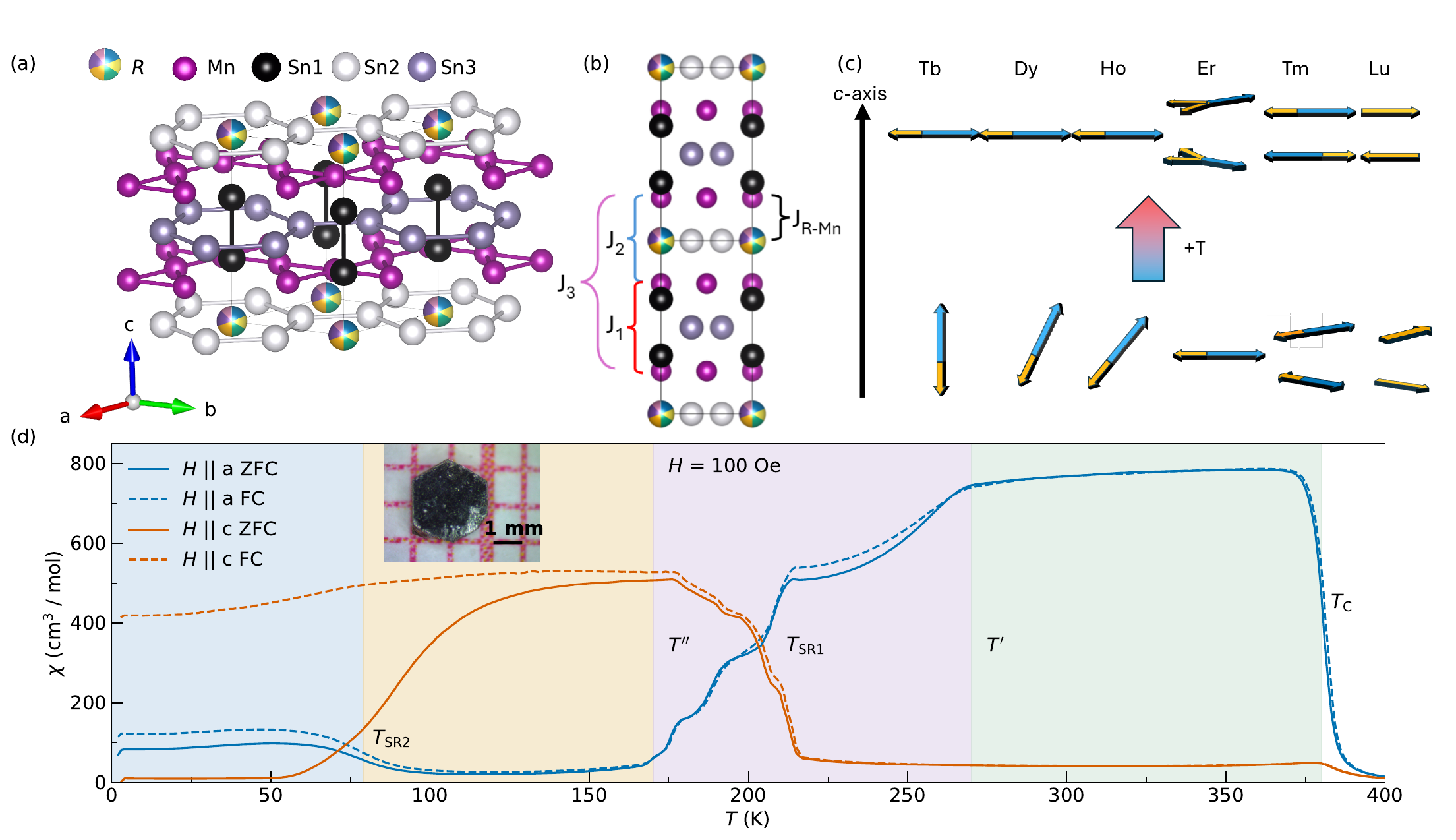}
  \caption{\label{fig:INTRO}
    (a) The crystal structure of \emph{HE}166 (same as other \ch{$R$Mn6Sn6}) showing the mixed rare-earth site and the kagome Mn layer. 
    (b) Magnetic couplings in the \ch{$R$Mn6Sn6} structure. 
    (c) Magnetic ground states of various \ch{$R$Mn6Sn6} materials in their pure form containing the rare-earth elements that were used to produce \emph{HE}166. The blue arrows represent the rare-earth moment of a given unit cell and the orange arrows represent the Mn moments in a unit cell. 
    (d) DC susceptibility vs. temperature measured in \emph{HE}166. Inset: Example image of a hexagonal \emph{HE}166 crystal on millimeter paper.
  }
\end{figure*}

The \ch{$R$Mn6Sn6} family is one such group of magnetic kagome systems. 
This material family crystallizes in the $P6/mmm$ space group and adopts a \ch{HfFe6Ge6} structure with a 3D layered-like structure stacked along the $c$-axis~\cite{venturini_magnetic_1991}. 
As shown in Fig.~\ref{fig:INTRO}a, the bottom layer of the unit cell is made of an $R$ triangular lattice interlaced with a honeycomb net of Sn atoms called the Sn2 site. 
Moving up the $c$-axis, there is a kagome Mn layer slightly offset from a triangular Sn1 layer, followed by a pure Sn3 honeycomb sublattice at the midpoint of the $c$-axis.
The neighboring Sn1 layers form Sn1-Sn1 dimers interpenetrating the Sn3 honeycomb net.
Those layers are then mirrored across the $ab$ plane to complete the unit cell. 

The magnetism in these materials depends on the interplay of a few magnetic couplings.
Both the Mn and $R$ layers have a ferromagnetic (FM) intra-layer coupling, while the coupling between the Mn and $R$ layers is generally antiferromagnetic (AFM). 
The overall magnetic ordering is then determined by the competition between the anisotropy of the $R$ ions and the inter-layer coupling of the Mn ions.
The Mn-Mn interactions can be described using four inter-layer Mn couplings: $J_1$, an FM coupling across the pure Sn layers; $J_2$, an AFM coupling across the \ch{$R$Sn2} layer; $J_3$, an FM next-nearest-neighbor coupling; and $J_{R-\text{Mn}}$, the coupling between the rare-earth and Mn ions that can be integrated out of the model to contribute an FM component to $J_2$ as illustrated in Fig.~\ref{fig:INTRO}b~\cite{lee_interplay_2023, ghimire_competing_2020}. 
Depending on the relative strength of the Mn-Mn couplings, the material can be FM, FiM, or AFM, while the rare-earth ion determines the direction of magnetic anisotropy.

The interplay among these couplings combined with the hybridization between local $f$-moments and itinerant $p$- and $d$-electrons leads to a variety of magnetic states depending on the choice of rare-earth.
In \ch{TbMn6Sn6} the magnetic ordering is ferrimagnetic (FiM), transitioning from a ground-state easy-axis to a higher-temperature easy-plane ordering~\cite{el_idrissi_magnetic_1991,yin_quantum-limit_2020, li_discovery_2023}. 
\ch{DyMn6Sn6} and \ch{HoMn6Sn6} have similar magnetic behavior, being in-plane FiM at higher temperatures, but having canted moments at lower temperatures~\cite{chen_magnetic_2021, kabir_unusual_2022}.
\ch{ErMn6Sn6} has an in-plane FiM ordering at low temperatures, and a spiral AFM ordering at higher temperatures with a large topological Hall effect (THE) arising from chiral fluctuations with the application of a field~\cite{fruhling_topological_2024, riberolles_new_2024, dhakal_anisotropically_2021}.
\ch{TmMn6Sn6} has a triple-spiral AFM ordering~\cite{liu_nontrivial_2023}. 
Materials with nonmagnetic rare-earth elements, \ch{LuMn6Sn6} and \ch{YMn6Sn6}, have spiral AFM ground state orders, with \ch{YMn6Sn6} showing a large THE over a wide temperature range due to a field-induced transverse conical spiral spin structure~\cite{venturini_magnetic_1991, ghimire_competing_2020}. 
The magnetic ground states of the \ch{$R$Mn6Sn6} materials with the rare-earth atoms relevant to this study are summarized in Fig.~\ref{fig:INTRO}c.

The varied magnetic behavior of \ch{$R$Mn6Sn6}  motivates investigating which behaviors are dominant and what unique behaviors arise in a high-entropy mix of rare earth elements. 
Here, we mix Tb, Dy, Ho, Er, Tm, and Lu to create a high-entropy form, hereafter referred to as \emph{HE}166.

Near zero fields, \emph{HE}166 shows a high-temperature ($\sim$380 K) in-plane FiM order which transitions into an out-of-plane FiM state before going through another spin reorientation for a canted magnetic ground state (Fig.~\ref{fig:INTRO}D).
This behavior is largely consistent with previous high-entropy studies of this family; however, those prior studies used different rare-earth mixes which, in their mono-rare-earth forms, all have FiM ground states.
Here we use a combination of six rare-earth elements, including three that show spiral AFM behavior, and unlike in previous studies, we observe a wider in-plane to out-of-plane transition showing a low-field incommensurate modulated magnetic state~\cite{liu_magnetic_2026, fruhling_characterization_2024, min_topological_2022}. 

In this work, we investigate the four distinct magnetic regimes in our \emph{HE}166 compound using magneto-transport and neutron diffraction measurements. 
These measurements reveal that high-entropy alloys can exhibit magnetic phenomena unique to any of the component non-mixed materials at different temperature and field regimes.

\section{\label{sec:methods}Methods}

\subsection{\label{sec:growth}Crystal Growth}

\emph{HE}166 single crystals were grown using a self-flux method similar to that of Refs.~\cite{venturini_magnetic_1991,clatterbuck_magnetic_1999}, but with one improvement.
Instead of using standard alumina crucibles, we used sealed tantalum tubes, which we found to yield larger crystals with less rare-earth deficiency. 
Tb, Dy, Ho, Er, Tm, and Lu pieces (99.9\%), Mn granules (99.98\%), and Sn pieces (99.999\%) were combined in a ratio ($R$Mn$_6$)$_{4.5}$Sn$_{95.5}$ with an equiatomic mixture of the various rare-earth elements. 
The reactants were placed in a tantalum crucible with a tantalum cap, which was sealed using an arc melting furnace under a small argon pressure.
The tantalum tube was then placed in an evacuated fused silica tube to protect it from oxygen, this silica tube was backfilled with argon to 1/3 atm with quartz wool padding below and above the tantalum crucible providing mechanical stability to the quartz tube.
The reactants were then heated in a box furnace to 1220$^\circ$ C at a rate of 100$^\circ$ C/h, held there for 24 h, cooled at a rate of 3$^\circ$ C/h to 575$^\circ$ C, annealed at this temperature for 24 h, and then decanted to remove excess flux.
Upon opening the ampule we noted no devitrification of the quartz itself and that the tantalum was pristine, maintaining its luster and mechanical strength, consistent with effective protection from oxidation.
Large millimeter-scale single crystals were produced as shown in the inset of Fig.~\ref{fig:INTRO}d.

\subsection{\label{sec:characterization}Characterizations}

The crystal structure and phase purity were verified using powder x-ray diffraction (PXRD) performed with a Bruker D8 ECO instrument in
the Bragg-Brentano geometry, using a copper source (CuK$\alpha$) and a LYNXEYE XE 1D energy-dispersive detector, and compared to \ch{ErMn6Sn6} (Fig. S1). 

Careful determination of the crystal structure of \emph{HE}166 was also done via single-crystal x-ray diffraction. 
Single crystal x-ray diffraction data were collected using a Bruker D8 Quest Kappa single crystal x-ray diffractometer equipped with an I$\mu$S microfocus source (Mo K$\alpha$ radiation, $\lambda$ = 0.71073 \AA) and PHOTON III CPAD area detector. 
Raw data frames were integrated with the Bruker SAINT program, and multiscan absorption corrections were applied using SADABS~\cite{krause_comparison_2015}. 
Preliminary models were generated via intrinsic phasing in SHELXT,  followed by additional refinement in SHELXL~\cite{sheldrick_shelxt_2015, sheldrick_crystal_2015}.
Detailed refinement results can be found in the supplemental Tables~S1-S4.
Occupancies of the rare-earth elements on the high-entropy site were constrained to values obtained from energy-dispersive x-ray spectroscopy data.
The VESTA program was then used for crystal visualizations~\cite{momma_vesta_2011}. 

The material's stoichiometry was determined using energy-dispersive x-ray spectroscopy (EDX) in an FEI Scios DualBeam electron microscope equipped with an Oxford detector (Fig.~S2), which was also used to image the distribution of the constituent elements (Figs. S3 and S4).
Magnetization characterization of samples was done using a Quantum Design MPMS3 with the sample mounted on a low-background quartz holder. Electrical transport was characterized in a 4-probe configuration using a Quantum Design PPMS Dynacool. 

\subsection{\label{sec:neutron_diffraction}Neutron Scattering}

Small-angle neutron scattering (SANS) and single-crystal neutron diffraction experiments were performed at the SANS-I and ZEBRA beamlines, respectively, both located at the Paul Scherrer Institute (PSI). 
The same \emph{HE}166 crystal was used for both experiments. The sample was a flat plate with dimensions of approximately $4 \times 2.5 \times 0.9$~$mm^3$, a mass of $\sim$ 50 mg, and the $c$-axis oriented normal to the plate.

For the SANS experiment, the sample was mounted inside an 11~T cryomagnet with a temperature range of 10 to 450~K. 
A longitudinal field was applied either parallel or perpendicular to the incoming neutron beam direction. SANS data were collected using two different sample orientations: (1) the $b^*-c^*$ plane within the 2D detector with the field applied parallel to the neutron beam, so along the real-space $a$-axis direction, and (2) the basal plane within the 2D detector with the real-space $c$-axis parallel to the beam and the field applied along the real-space $a$-axis. 
For both sample orientations, data were acquired with 4.8$~\angstrom$ neutrons using a high-Q configuration and a low-Q configuration, with sample-to-detector distance of 6~m and 18~m, respectively. 
At each field and temperature step, rocking scans were performed over a range of 9$\degree$ in steps of 0.25~$\degree$. 

We performed both zero-field and in-field ZEBRA experiments.
For the zero-field experiment, the \emph{HE}166 sample was mounted inside a closed-cycle refrigerator (CCR) spanning a temperature range of 6 to 300~K. 
The crystal was mounted on a 4-circle goniometer and oriented with the $c$-axis perpendicular to the neutron beam when $\chi=0$. 
The diffraction data for the temperature-dependent order parameter near $\bf{Q}$~=~(002) was collected using 2.305~\angstrom~neutrons and an $80^{\prime\prime}$ collimation. 
We then collected a series of $\omega$-scans on various Bragg peak positions to extract the structure factors required to refine the magnetic structure of \emph{HE}166 at 300~K, 150~K, and 6~K. 
For the field experiment, the \emph{HE}166 crystal was mounted in a 6~T vertical magnet with the $a$-axis parallel to the field. 
In-field order parameters and $\omega$-scans were collected using 1.383~$\angstrom$ incident neutrons.

\section{\label{sec:results}Results and Discussion}

\subsection{Crystal Structure}

Several large hexagonal samples were grown as shown in the inset of Fig.~\ref{fig:INTRO}d. 
According to EDX, the samples had an averaged stoichiometry of \ch{Tb_{0.18}Dy_{0.15}Ho_{0.17}Er_{0.19}Tm_{0.17}Lu_{0.12}Mn_{6.2}Sn_{6.1}} and they were found using XRD to crystallize in the $P6/mmm$ space group common to this family of materials.
The structural parameters were calculated from single-crystal XRD to be $a = 5.5173(6)$ \angstrom~and $c = 9.0053(11)$ \angstrom.

\subsection{Magnetization}

We used DC magnetic susceptibility to probe different magnetic phases at a small field of 100 Oe in \emph{HE}166, as shown in Fig.~\ref{fig:INTRO}d. 
Data were collected with the field both along the $a$-axis (in the hexagonal crystal plane) and along the $c$-axis (perpendicular to the plane). 
Both data sets show the onset of FiM ordering at \TC~= 380(6) K and a transition from an easy-plane to easy-axis anisotropy at $T_{\text{SR}1}$ = 207(3) K, followed by a spin reorientation into a canted state at $T_{\text{SR}2}$ = 79(11) K.
The transition temperatures were determined from the peaks in the derivative of the ZFC $H||a$ data (Fig.~S5). 

This behavior is qualitatively similar to previous high-entropy \ch{$R$Mn6Sn6} reports, which also found a FiM easy-plane order at high temperatures with two lower-temperature spin reorientations~\cite{liu_magnetic_2026, fruhling_characterization_2024, min_topological_2022}. 
However, unlike the previous reports, we observe a broader transition from in-plane to out-of-plane FiM order with multiple steps in our sample.
We treat this transitional region as a separate region of interest that starts at $T'$ = 270(10) K and ends at $T''$ = 170(10) K (purple area in Fig.~\ref{fig:INTRO}d). 

\begin{figure}
\centering
  \includegraphics[width=.48\textwidth]{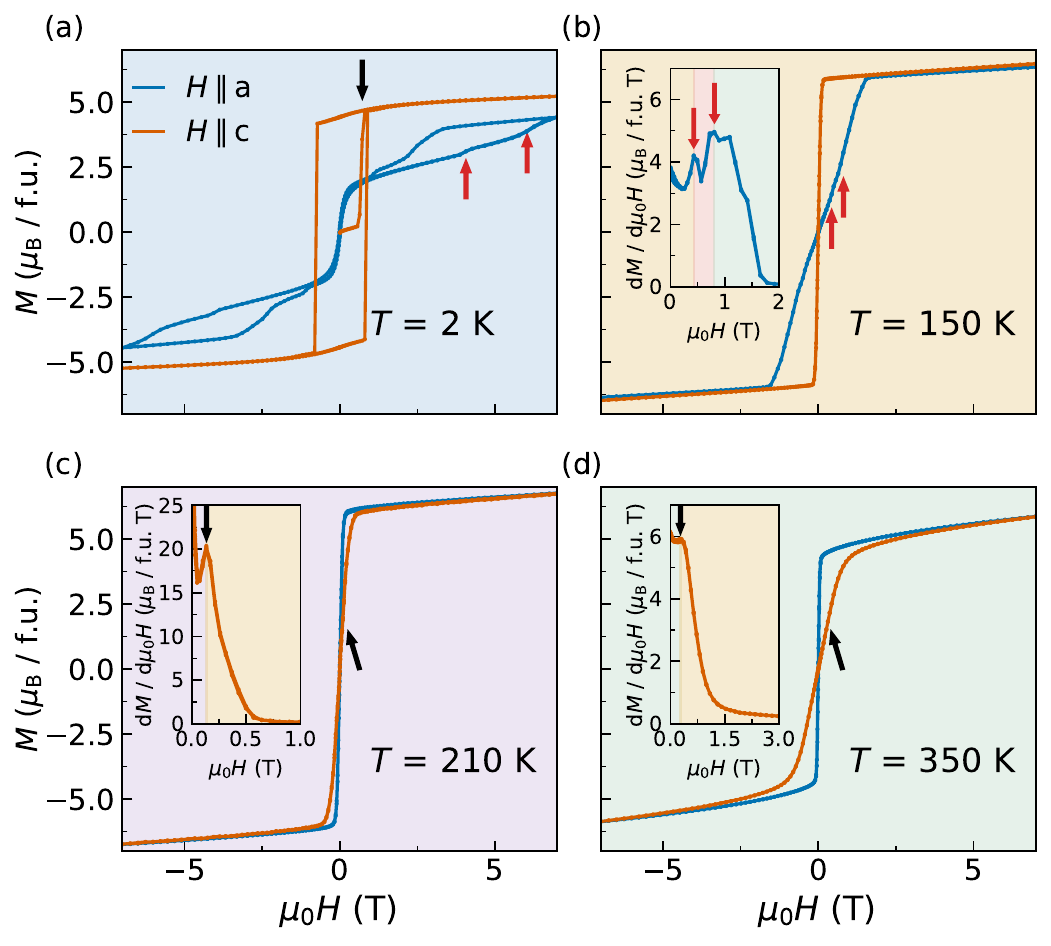}
  \caption{\label{fig:M(H)}
    Magnetization as a function of field measured in \emph{HE}166 at (a) 2 K, (b) 150 K (inset: derivative of magnetization with respect to field with $H\|a$), (c) 210 K (inset: derivative of magnetization with respect to field with $H\|c$), and (d) 350 K (inset: derivative of magnetization with respect to field with $H\|c$).
    The background color of each panel corresponds to the temperature regimes in Fig.~\ref{fig:INTRO}d.
    Arrows point to critical fields used to construct the phase diagrams in Fig.~\ref{fig:PHASES}.
    In the insets orange backgrounds represent easy-axis and axis-polarized regions, red represents the metamagnetic region in between the field-induced transitions present in the in-plane field data, and green represents in-plane polarized region.
  }
\end{figure}

The four regions in Fig.~\ref{fig:INTRO}d were characterized by measuring magnetization as a function of field. 
Fig.~\ref{fig:M(H)} shows the magnetization curves at representative temperatures within each region with field applied along the $a$-axis and along the $c$-axis.
At low temperatures, in the canted FiM region (blue), two metamagnetic transitions are observed when the field is applied along the $a$-axis.
At 2 K, these transitions occur at 4.1(2) T and 6.2(2) T, marked by the red arrows, when the field is increased from zero after the sample has been zero-field cooled.
A significant hysteresis is seen at higher fields.
Note that the in-plane magnetization does not reach saturation at 7~T, but the out-of-plane one does.
When the field is applied along the $c$-axis, magnetization saturates by 0.9 T, and a typical hysteresis loop is observed with a coercive field of 0.8~T.
While there are no clear metamagnetic transitions when the field is increased or decreased from a point of saturation, when the sample is zero-field cooled, there is a characteristic field at which the magnetization sharply increases, marked by the black arrow.
This may be due to the alignment of domains, or a metamagnetic transition from the canted state to $c$-axis polarization.

In the easy-axis region (orange) between 79 and 170 K, the FiM state is much softer, with negligible hysteresis when the field is along either the $a$-axis or the $c$-axis.
When $H||a$, there are still two metamagnetic transitions, red arrows, although they are more subtle and can be more easily observed as peaks in the derivative of the magnetization with respect to field as shown in the inset of Fig.~\ref{fig:M(H)}b.
These transitions are also shifted to lower fields (0.4(1)T and 0.8(1)T at 150 K).
When $H||c$, there are no field-induced transitions, and the magnetization saturates at low fields.

Within the spin-reorientation region (purple), as the temperature increases, the easy and hard field directions flip; however, the magnetization shows similar soft FiM behavior, when the field is along either direction.
A subtle metamagnetic transition, black arrow, appears in the out-of-plane magnetization, as the easy-plane behavior begins to emerge, and the in-plane metamagnetic transitions are suppressed near $T_{\text{SR1}}$.
The inset of Fig.~\ref{fig:M(H)}c shows the derivative of magnetization with respect to field, showing the low-field metamagnetic transition as a low-field peak.
At 210 K, this transition occurs at 0.14(4) T.
The only other change is a shift in the saturation fields as the temperature varies throughout the region. 

In the easy-plane temperature region (green), the field does not induce any transitions when applied along either the $a$-axis or $c$-axis.
The two magnetization curves differ only in their saturation fields, as expected for a simple FiM easy-plane system; the spins polarize more easily when the field is in the plane compared to out of plane.
In order to demarcate the easy-plane region from the axis-polarized region, the field where the derivative of the magnetization begins to rapidly decrease is used, black arrow in Fig.\ref{fig:M(H)}d. 
Qualitatively, this behavior is similar to the magnetization behavior previously reported in high-entropy forms of \ch{$R$Mn6Sn6}, with changes only noticeable in the magnitude of the hysteresis, metamagnetic transition fields, and transition temperatures. 
That is, the inclusion of rare-earth elements that have AFM ordering in their pure \ch{$R$Mn6Sn6} form, does not seem to alter the fundamental behavior of the high-entropy form's bulk magnetization. 

\begin{figure*}
\centering
  \includegraphics[width=\textwidth]{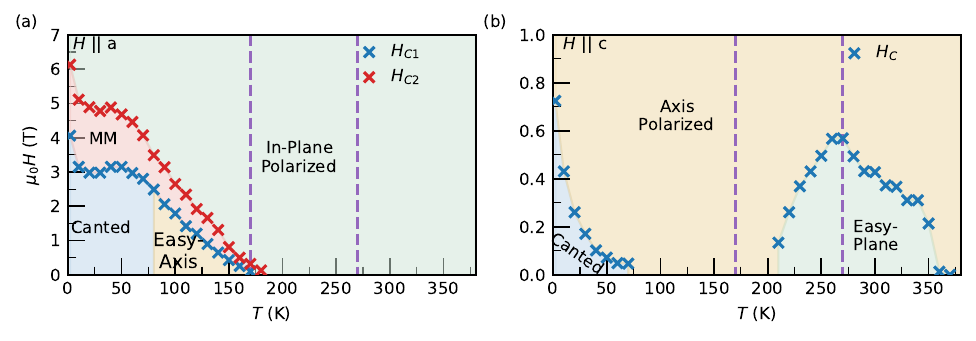}
  \caption{\label{fig:PHASES}
    (a) Magnetic phase diagram of $HE$166 when the field is applied along the $a$-axis. (b) Phase diagram with field applied along the $c$-axis. Labeled regions correspond to canted, in-plane, and axis spin orientations as shown in Fig.~\ref{fig:INTRO}d with the additional red color area representing the metamagnetic (MM) region between the two field-induced transitions in the $H\parallel a$ $M(H)$ data. Purple dashed lines correspond to the edges of the transitional region.
  }
\end{figure*}

The field-induced transitions were determined from peaks in the derivative of magnetization with respect to field (see Fig.~S6). 
The evolution of the magnetization in each region was then tracked in supplemental Fig.~S7. 
From these magnetization curves, the in-field magnetic phase diagrams for $H\|a$ and $H\|c$ were constructed and shown in Fig.~\ref{fig:PHASES}.
The red color in Fig.~\ref{fig:PHASES}a is the metamagnetic (MM) region between the two field-induced transitions present when field is applied in-plane. 
As the transitional region does not correspond to any clear qualitative behavior changes in the $M(H)$ curves, its borders are marked by the purple dotted lines.

\subsection{Electrical Transport}

\begin{figure}
\centering
  \includegraphics[width=.48\textwidth]{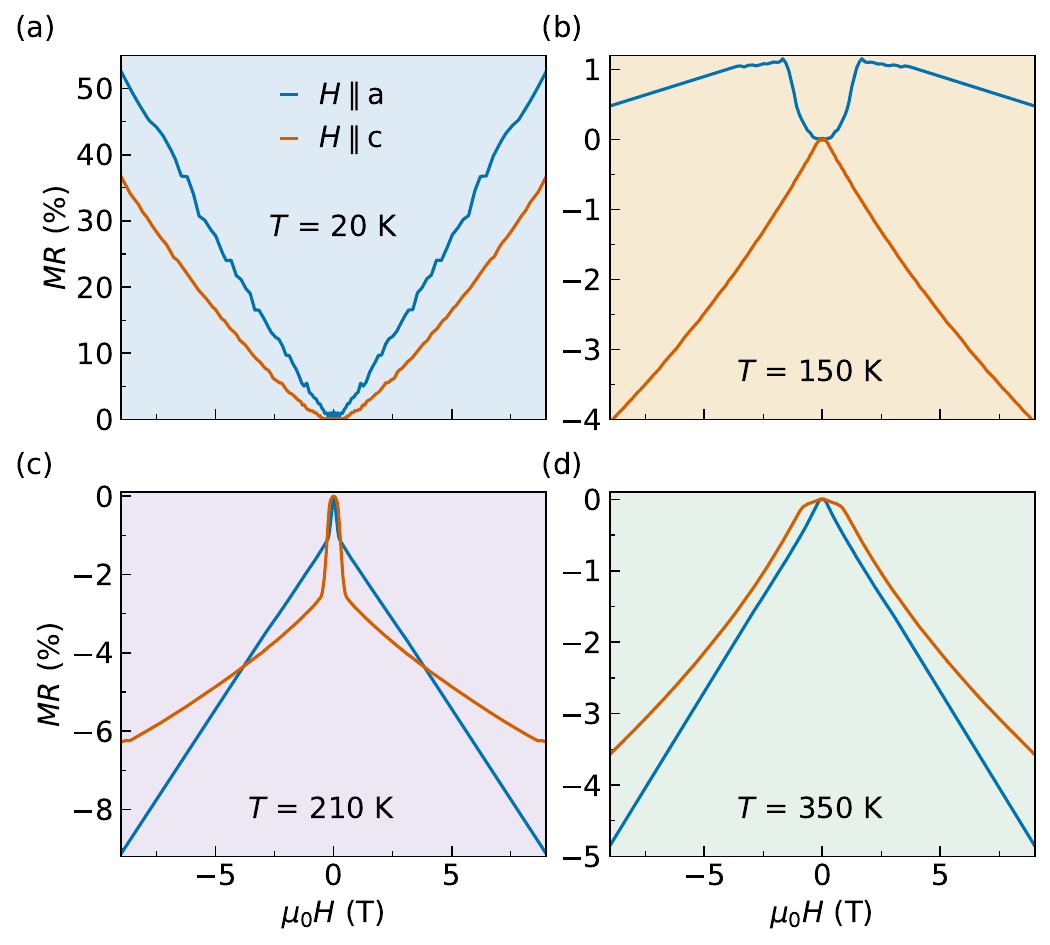}
  \caption{\label{fig:MR(H)}
    Longitudinal magnetoresistance as a function of field for \emph{HE}166 at (a) 20 K, (b) 150 K, (c) 210 K, and (d) 350 K.
    The background color of each panel corresponds to the temperature regimes in Fig.~\ref{fig:INTRO}d.
    All four panels show in-plane resistivity data, measured under both in-plane field (blue line) and out-of-plane field (orange line). 
  }
\end{figure}

A recent study of the high-entropy form \ch{(Gd,Tb,Dy,Ho,Er)1Mn6Sn6} has revealed a monotonically increasing magnetoresistance (\emph{MR}) with linear field dependence of resistivity at all temperatures~\cite{liu_magnetic_2026}.
The pure forms of \ch{$R$Mn6Sn6} made with $R=$ Gd, Tb, Dy, and Ho exhibit collinear FiM orders. 
Only \ch{ErMn6Sn6} supports a noncollinear order, and then, only at higher temperatures.~\cite{riberolles_new_2024}.
In this work, we chose rare-earth elements that supported noncollinear ordering, namely Er, Lu, and Tm~\cite{venturini_magnetic_1991}.
This choice has led to a nonmonotonic \emph{MR} behavior in our \emph{HE}166, \ch{(Tb,Dy,Ho,Er,Tm,Lu)1Mn6Sn6}, unlike the monotonic behavior reported in Ref.~\cite{liu_magnetic_2026}.

The field dependence of \emph{MR} = $\left[R(H)-R(0)\right]/R(0)\times 100\%$ is presented in Fig.~\ref{fig:MR(H)} in the four temperature regimes derived from the magnetization data in Fig.~\ref{fig:INTRO}d.
Notably, a monotonically increasing linear field dependence of \emph{MR} is observed only at the lowest temperature $T=20$~K (blue region), while a nonmonotonic or decreasing \emph{MR} behavior is observed in the other three temperature regimes. 

In the easy-axis region (orange), when the field is applied along the $a$-axis, the \emph{MR} shows a reversal from increasing to decreasing at 0.9~T, where magnetization saturates (Fig.~\ref{fig:M(H)}b), and the spins align with the applied field.
However, when the field is applied along the $c$-axis the \emph{MR} monotonically decreases with increasing field.
This behavior is expected as the spins of a system polarize and align with the field, decreasing the scattering of charge carriers due to spin-disorder~\cite{vitayaya_magnetoresistance_2024}.

In the transitional region (purple), where the system transitions from easy-axis to easy-plane, both field directions show a negative \emph{MR}; however, in each case there is a notable change of slope at a field corresponding to magnetic saturation, namely 0.24~T for $H\|a$ and 0.51~T for $H\|c$ (Fig.~\ref{fig:M(H)}c). 
The origin of this change is explored further in Sec.~\ref{sec:Neutron_Diffraction_2}.

Finally, in the high-temperature easy-plane region (green), \emph{MR} follows the expected behavior of a FiM, with a change of slope upon magnetization saturation and a linear field dependence at higher fields.

For completeness, we show the temperature dependence of resistivity at different fields applied along the $a$-axis and $c$-aixs in Fig.~S8.
Two interesting findings in that figure are (1) a relatively large residual resistivity ratio (RRR=98) for a high-entropy form, and (2) resistivity anomalies at the onset of magnetic ordering and the higher-temperature spin-reorientation at the same temperatures as those determined by magnetization.
Hall resistivity measurements (Figs.~S9-S11) reveal that the transverse resistivity reflects the metamagnetic transitions at higher temperatures where the anomalous Hall effect becomes notable. 
The large anomalous Hall effect which dominates at higher temperatures is in line with the results of Ref.~\cite{liu_magnetic_2026}. 
No additional Hall-resistivity feature was resolved across the transitional region. 
The principal qualitative change occurs near $T_{\text{SR1}}$, where the field-induced jump in resistivity shifts from occurring in the $H\parallel a$ data to the $H \parallel c$ data upon warming, consistent with the corresponding change in the field-dependent magnetization.

\subsection{\label{sec:Neutron_Diffraction_1}Neutron Diffraction at Zero Field}

\begin{figure}
\centering
  \includegraphics[width=.48\textwidth]{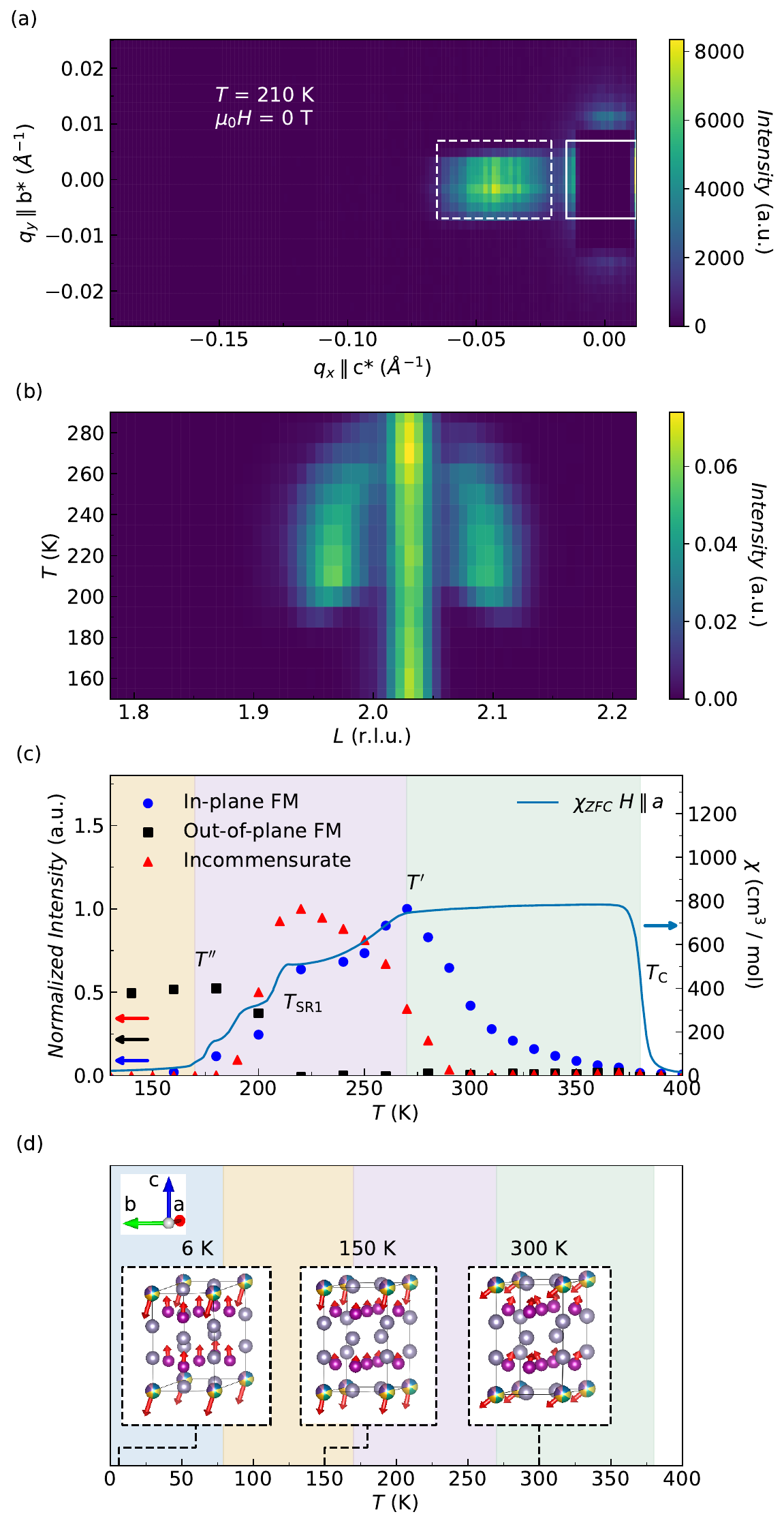}
  \caption{\label{fig:ZERO_FIELD_NEUTRON}
    (a) SANS intensity at 210 K and zero-field. The white solid box highlights the region near the beam stop integrated in order to estimate the ferromagnetic in-plane contribution, and the white dotted box highlights the incommensurate peak. 
    (b) Neutron high-angle diffraction intensity in zero-field around the (002) nuclear peak. Incommensurate satellite peaks appear between 170 K and 270 K. 
    (c) Left axis: the relative intensity of three different scattering processes; the in-plane FM contribution of the Q$\sim$0 integration region (blue circles) as shown in the solid box in (a), its out-of-plane contribution (black squares), and the intensity of the satellite peak (red triangles) shown in the dotted box in (a). Right axis: the DC ZFC susceptibility with field along the $a$-axis recreated from \ref{fig:INTRO}a to indicate transition temperatures.
    (d) Neutron refinement structures from high-angle neutron scattering at 6 K, 150 K and 300 K. The structures visualize the relative in-plane and out-of-plane components obtained from the $U(1)$ basis-vector refinement described in the text and SM.
  }
\end{figure}

We next characterized the collective magnetism of \emph{HE}166 using neutron scattering techniques. 
Both SANS (Fig.~\ref{fig:ZERO_FIELD_NEUTRON}a) and single-crystal high-angle neutron diffraction (Fig.~\ref{fig:ZERO_FIELD_NEUTRON}b) detected commensurate as well as incommensurate Bragg peaks, respectively indexed by a propagation vector $k_\text{com}=(0,0,0)$ and $k_\text{inc}=(0,0,\delta)$. 
The presence of $k_\text{inc}$ Bragg peaks indicates an out-of-plane spin modulation, reminiscent of the spin spirals observed in the single-ion analogue compounds, such as ErMn$_6$Sn$_6$, YMn$_6$Sn$_6$, and LuMn$_6$Sn$_6$ \cite{fruhling_topological_2024, venturini_magnetic_1991, ghimire_competing_2020}. 

We first isolated the $k_\text{inc}$ component by integrating over the SANS subspace where the incommensurate peaks arise (see the dashed white box in Fig.~\ref{fig:ZERO_FIELD_NEUTRON}a). 
For the $k_\text{com}=(0,0,0)$ component, represented by the blue data points in Fig.~\ref{fig:ZERO_FIELD_NEUTRON}c, we integrated the SANS intensity near the beam stop ($|\bf{Q}|\sim0$) using the solid white box (see Fig.~\ref{fig:ZERO_FIELD_NEUTRON}a) oriented such that the integration is performed mostly along the (00$L$) direction, thereby providing an estimate of the in-plane ferromagnetism, though other magnetic inhomogeneities could also contribute to the $|\vec{Q}| \sim 0$ scattering.
As shown in Fig.~\ref{fig:ZERO_FIELD_NEUTRON}c, the temperature dependence reveals that these in-plane ferromagnetic correlations onset near $T_\text{C}=380$~K and grow smoothly upon cooling.
Below $T'=270$~K, a gradual shift in spectral weight occurs, suppressing the in-plane ferromagnetism as intensity builds at the incommensurate Bragg position (red data points).
This gradual shift of intensity occurs down to $T_\text{SR1}=207$~K where the intensity of both $k_\text{inc}$ and in-plane ferromagnetism begin to vanish.
Upon cooling, the in-plane ferromagnetic correlations can be seen to drop in intensity similarly to the in-plane susceptibility, however, the temperature sampling of the neutron measurements (every 10 K) does not allow for the individual steps in the susceptibility to be unambiguously associated with corresponding changes in the magnetic scattering.

To understand the shift of spectral weight occurring below $T_\text{SR1}=207$~K, we also integrated the $|\vec{Q}\sim0|$ scattering contribution using a SANS geometry that collects scattering entirely within the basal (HK0) plane (SM, Fig.~S13e), making it sensitive to out-of-plane ferromagnetism. 
This integration is represented by the black data points in Fig.~\ref{fig:ZERO_FIELD_NEUTRON}c and clearly shows an onset below $T_\text{SR1}=207$~K, which is precisely where both the in-plane ferromagnetism and the incommensurate Bragg contributions begin to disappear.

The temperature-dependent SANS, explained above, is consistent with the magnetization data for which an in-plane ferromagnetic state is observed below $T_\text{C}$, while an out-of-plane ferromagnetic state is developed below $T_\text{SR1}$. 
The incommensurate modulated state arises and falls within the transitional phase from easy-plane to easy-axis ferromagnetism. 
This is consistent with the diffraction data (Fig.\ref{fig:ZERO_FIELD_NEUTRON}b) where incommensurate satellite peaks are only observed between $T'=270$~K and $T''=170$~K.
The exact position of these satellite peaks is temperature-dependent with an incommensurate modulation of $\sim$25~nm at 270~K, which decreases slightly upon cooling to reach $\sim$11~nm at 180~K (see Fig.~S13).

To gain further insights into the zero-field temperature-dependent $\bf{k}=0$ spin structure changes of \emph{HE}166, we measured the structure factors of several magnetic and nuclear Bragg reflections at $T$ = 6, 150, and 300~K on the single crystal neutron diffractometer ZEBRA. 
Magnetic refinement details, including Lorentz, absorption, and extinction corrections, are provided in the SM Fig.~S12~\cite{rodriguez-carvajal_recent_1993,rodriguez-carvajal_crystallographic_2002,sheldrick_crystal_2015, bumps_software}.
For all temperatures, the data were refined using a $q=0$ magnetic model containing a coherent $U(1)$ in-plane Mn order parameter, a uniform Mn out-of-plane magnetization ($M^\text{Mn}_{c}$), and a canted effective $R$ moment with both out-of-plane ($M^R_c$) and in-plane ($M^R_{ab}$) components.
The $U(1)$ in-plane Mn order parameter is spanned by the 191.238 and 191.239 basis vectors, using the Bilbao magnetic-space-group labeling convention~\cite{perez-mato_symmetry-based_2015, aroyo_bilbao_2006}. 
These two basis vectors generate orthogonal components of the same 120$\degree$ spin arrangement on each Mn triangular plaquette, yielding zero net in-plane Mn magnetization. We parameterize the in-plane Mn component as $M^\text{Mn}_{ab}\left[\cos(\alpha)\psi_{238}+\sin(\alpha)\psi_{239}\right]$, where $\alpha$ selects the in-plane phase of the 120$\degree$ configuration.
The refined magnetic parameters for \emph{HE}166 are summarized in Table~\ref{tab:basis_vectors} at the three representative temperatures.

\begin{table}[b] 
\setlength{\tabcolsep}{3.5pt} 
\small 
\begin{tabular}{cccccccc}
\hline\hline
\textbf{Temp.} & $\alpha$ & $M^{\text{Mn}}_{ab}$ & $M^{\text{Mn}}_{c}$ & $M^{R}_{ab}$ & $M^{R}_{c}$ & $\theta^{R}$ & $R_{\rm abs}$ \\
(K) & ($\degree$) & ($\mu_B$) & ($\mu_B$) & ($\mu_B$) & ($\mu_B$) & ($\degree$) & \\
\hline
300  & 213.2 & 1.22 & 2.06 & 5.04 & -2.88  & 60.2 & 0.286 \\
150  & 0.09  & 0.06 & 0.69 & 3.02 & -9.21  & 18.1 & 0.285 \\
6    & 0.005 & 0.11 & 1.36 & 5.89 & -11.37 & 27.4 & 0.294 \\
\hline\hline
\end{tabular}
\caption{Magnetic parameters obtained from the coherent $U(1)$ basis-vector refinement of \emph{HE}166. The angle $\alpha$ defines the in-plane Mn basis-vector phase $M^\text{Mn}_{ab}\left[\cos(\alpha)\psi_{238}+\sin(\alpha)\psi_{239}\right]$. The angle $\theta^{R}$ is the canting angle of the effective $R$ moment away from the dominant $c$-axis direction.}
\label{tab:basis_vectors}
\end{table}

The magnetic refinement reveals a strong change between 300~K and the lower temperatures.
This change in magnetic structure is visualized in Fig.~\ref{fig:ZERO_FIELD_NEUTRON}d.
At 300~K the refinement gives a sizable in-plane Mn component and a strongly canted effective $R$ moment. At 150~K and 6~K, the fitted in-plane Mn component is very small, while the $R$ moment is mostly oriented along the $c$-axis with a finite in-plane canting component.
The sharp susceptibility drop near $T''$ (Fig.~\ref{fig:INTRO}d) makes this change plausible.
We note that we describe the orientation and amplitude of the $R$ moment as an average over the entire sample, which is likely varying among all different $R$ ions forming \emph{HE}166.
We also note that the representations in Fig.~\ref{fig:ZERO_FIELD_NEUTRON}d are intended to illustrate the relative magnitudes of the Mn and $R$ moments and the relative in-plane and out-of-plane components.

\subsection{\label{sec:Neutron_Diffraction_2}Neutron Diffraction in Field}

\begin{figure}
\centering
  \includegraphics[width=.48\textwidth]{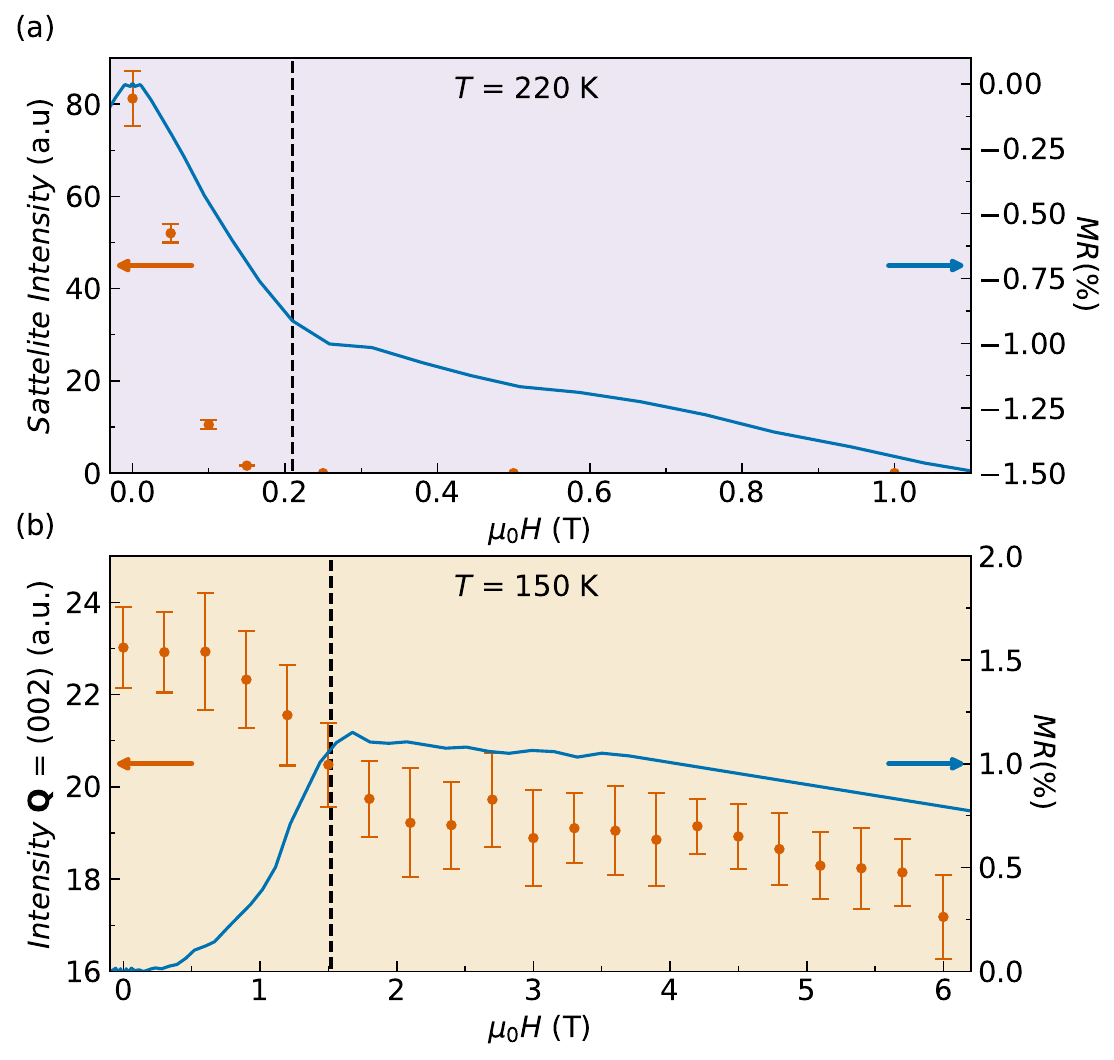}
  \caption{\label{fig:IN_FIELD_NEUTRON}
    (a) Left axis: Satellite peak intensity at 220 K as a function of field showing complete suppression by 0.2 T. Right axis: Magnetoresistance at 220 K showing a change of slope at the same field where the satellite intensity is suppressed.
    (b) Left axis: $\mathbf{Q}$ = (002) peak intensity vs. field at 150 K showing a kink and sharp decrease in intensity near 1.5 T. Right axis: Magnetoresistance at 150 K showing a corresponding change from increasing to decreasing magnetoresistance.
  }
\end{figure}

We also performed in-field neutron scattering experiments, as our transport measurements clearly reveal distinct \emph{MR} changes between the zero-field limit and the in-field saturated magnetized state of \emph{HE}166. 
The most prominent \emph{MR} variations occur between $T'=270$~K and $T''=170$~K (purple region) as well as between $T_{\text{SR2}}=79$~K and $T''$ (orange region), which respectively correspond to the incommensurate magnetized state of \emph{HE}166 and the out-of-plane FiM state. Consequently, we qualitatively explore the in-field neutron diffraction behavior at $T = 220\text{ K}$ and $T = 150\text{ K}$ to investigate both regimes.

The \emph{MR} and the neutron scattering data for $T=220$~K are both compared in Fig.~\ref{fig:IN_FIELD_NEUTRON}a.
The in-field SANS intensity, which we integrated over the incommensurate peak (see dashed white line in Fig.~\ref{fig:ZERO_FIELD_NEUTRON}a), clearly shows that the incommensurate magnetism disappears above $\sim$0.2~T. 
The suppression of the incommensurate modulation occurs at the same characteristic field scale as the change in the slope of the \emph{MR} indicating coupling between the modulated magnetism and electronic transport.
Notably, the negative slope of the \emph{MR} is steepest during the rapid suppression of the incommensurate state.
This behavior could potentially be understood as arising as a long-wavelength non-collinear state forces conduction electrons to continuously realign their spin directions, thereby enhancing the spin-flip scattering rate. 
As the field aligns the spins in this situation, the spin-flip scattering rate would rapidly decrease.
Furthermore, the incommensurate order may also modify the Fermi surface via gap openings produced by a magnetic zone folding, which is not present in the $\bf{k}$=0 state.

An in-plane field also induces a slope change in the \emph{MR} at $T$=150~K near $H$=1.5~T.
We compare this \emph{MR} behavior with the field-dependence of the $\bf{Q}$=(002) Bragg peak in Fig.~\ref{fig:IN_FIELD_NEUTRON}b. 
The field-dependent intensity of the (002) peak also displays a clear kink at $H$=1.5~T, which is indicative of a spin reorientation during the transition towards the field-polarized state.
The magnetic structure factor for (002) probes the in-plane antiferromagnetic components between the $R$ ion and the Mn ions, such that it vanishes if both sublattices align with equal magnitude. 
Zero-field magnetic refinement at 150~K (Table~\ref{tab:basis_vectors}) reveals that the primary intensity contribution to the (002) peak originates from the in-plane component of the $R$ ions. 
Under an applied field $H\parallel a$, the spin reorientation reduces the (002) intensity, indicating that the Mn spins rotate from an out-of-plane orientation to align along the in-plane directions.
Consequently, the field-polarized state suppresses the non-collinear spin component between the $R$ and Mn ions relative to the zero-field state. 
This change in the non-collinear component of the spin structure seems to couple with the existence of a unique magnetoresistive behavior that has not been observed in other high-entropy forms of \ch{$R$Mn6Sn6}.

\section{\label{sec:conclusion}Conclusion}

The high-entropy \ch{$R$Mn6Sn6} compound studied here shows a distinctive temperature dependence of magnetization and a nonmonotonic field dependence of magnetoresistance compared to previously reported high-entropy forms.
Neutron diffraction experiments reveal that this is due to a non-collinear or modulated spin component from the rare-earth moments.
While these intricate details of the magnetism cannot be detected by measuring the magnetization as a function of field, when the magnetoresistance is measured as a function of field, there are clear field induced changes in behavior that correlate to the suppression of these spin textures' contributions with field.


As discussed in previous studies, both the ground state and high-temperature behavior of \ch{$R$Mn6Sn6} compounds depend strongly on the choice of rare-earth element, which influences the Mn–Mn exchange coupling across the rare-earth layer as well as the magnetic anisotropy. 
Comparison with earlier reports indicates that several characteristics of high-entropy rare-earth variants are robust with respect to the specific rare-earth composition. 
In particular, high-entropy \ch{$R$Mn6Sn6} compounds generally exhibit a high-temperature easy-plane FiM phase that transitions to an intermediate-temperature easy-axis phase, followed by a low-temperature canted FiM state. 
These phases are often accompanied by metamagnetic transitions that give rise to intermediate field-induced magnetic states prior to full spin polarization for in-plane applied fields.

A unique characteristic of the \emph{HE}166 studied here is the wide transitional region between the easy-plane and easy-axis regions, not previously reported.
This behavior emphasizes the role of the rare-earth coupling in this family and the importance of choosing rare-earth elements which do not have a common ground state if one wishes to engineer high-entropy materials with unconventional magnetism and magnetotransport in the future.

\section*{Acknowledgments}
The work at Boston College (crystal growth, transport and magnetization measurements, and neutron scattering) was supported by the National Science Foundation under award number DMREF-2522383. The work at the Air Force Research Laboratory (crystal growth and characterization) was supported by the Air Force Office of Scientific Research LRIRs 23RXCOR003 and 26RXCOR010. Further support was provided by the AFRL-NSF INTERN Program. J.Y.C. and N.J.F. acknowledge Welch Foundation Grant AA-2056-20240404 and NSF-DMR-2505304 for funding. The support for neutron scattering was provided by the Center for High-Resolution Neutron Scattering, a partnership between the National Institute of Standards and Technology and the National Science Foundation under Agreement No. DMR-2010792. The identification of any commercial product or trade name does not imply endorsement or recommendation by the National Institute of Standards and Technology. The neutron scattering experiments were performed on
SANS-I and ZEBRA instruments at the Swiss spallation neutron source SINQ, Paul Scherrer Institute, Villigen, Switzerland.




\bibliography{Bibliography}

\end{document}



\title{Supplemental Material:\\A high-entropy form of \ch{$R$Mn6Sn6} with distinct magnetotransport regimes correlated to different magnetic structures}

\author{Kyle~W.~Fruhling$^{\ast\dagger}$}
\email{fruhling@bc.edu}
\affiliation{Department of Physics, Boston College, Chestnut Hill, MA 02467, USA}

\author{Jonathan~Gaudet}
\affiliation{NIST Center for Neutron Research, National Institute of Standards and Technology, Gaithersburg, MD 20899, USA}
\affiliation{Department of Materials Science and Eng., University of Maryland, College Park, MD 20742, USA}

\author{William~D.~Ratcliff}
\affiliation{NIST Center for Neutron Research, National Institute of Standards and Technology, Gaithersburg, MD 20899, USA}
\affiliation{Department of Physics and Department of Materials Science and Engineering, University of Maryland, College Park, MD 20742, USA}

\author{Jonathan~S.~White}
\affiliation{PSI Center for Neutron and Muon Sciences, Forschungsstrasse 111, 5232 Villigen, PSI, Switzerland}

\author{Siddharth~Nandanwar}
\affiliation{Department of Physics, Boston College, Chestnut Hill, MA 02467, USA}

\author{Noah~J.~Fau}
\affiliation{Department of Chemistry and Biochemistry, Baylor University, Waco, TX 76798, USA}

\author{Gregory~T.~McCandless}
\affiliation{Department of Chemistry and Biochemistry, Baylor University, Waco, TX 76798, USA}

\author{Enrique~O.~Gonz\'{a}lez-Delgado}
\affiliation{Department of Physics, Boston College, Chestnut Hill, MA 02467, USA}

\author{Julia~Y.~Chan}
\affiliation{Department of Chemistry and Biochemistry, Baylor University, Waco, TX 76798, USA}

\author{Oksana~Zaharko}
\affiliation{PSI Center for Neutron and Muon Sciences, Forschungsstrasse 111, 5232 Villigen, PSI, Switzerland}

\author{Michael~A.~Susner}
\affiliation{Foundational Technologies Directorate, Air Force Research Laboratory, Wright-Patterson Air Force Base, Ohio 45433, USA}

\author{Michael~J.~Newburger}
\affiliation{Foundational Technologies Directorate, Air Force Research Laboratory, Wright-Patterson Air Force Base, Ohio 45433, USA}

\author{Fazel~Tafti}
\email{fazel.tafti@bc.edu}
\affiliation{Department of Physics, Boston College, Chestnut Hill, MA 02467, USA}



\maketitle


\pagebreak

\section{\label{sec:Crystal Characterization}Crystal Characterization}

\subsection{\label{sec:PXRD}PXRD}

\begin{figure*}
\centering
  \includegraphics[width=.6\textwidth]{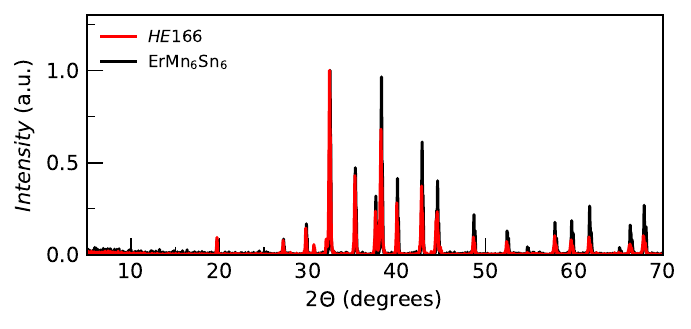}
  \caption{\label{fig:PXRD}
    PXRD scan of \emph{HE}166 with a scan of \ch{ErMn6Sn6}. The peaks at 30.6$^\circ$ and 32.0$^\circ$ correspond to excess Sn flux on the surface of the crystals.
  }
\end{figure*}

Powder x-ray diffraction (PXRD) was used to confirm the crystal structure and lack of impurity phases in the \emph{HE}166 crystals. 
Fig.~\ref{fig:PXRD} shows a comparison between a \emph{HE}166 PXRD scan and \ch{ErMn6Sn6} confirming the expected $P6/mmm$ space group.
Due to the change in lattice parameters from the difffering rare-earth sizes, the peaks are shifted slightly in \emph{HE}166 compared to \ch{ErMn6Sn6}.
There are also notable unmatched peaks at 30.6$^\circ$ and 32.0$^\circ$ due to excess Sn flux on the surface of the crystals.
No other impurity phases were found in PXRD.

\subsection{Single Crystal XRD}

Single crystal x-ray diffraction (XRD) was used to accurately determine the crystal parameters.
The relative occupancies of the various $R$ ions was constrained by EDX results.
The refinement results are summarized in Tables~\ref{tab:REFINEMENT}-\ref{tab:INTERATOMIC_DISTANCES}.

\begin{table*}[htbp] \centering \caption{\label{tab:REFINEMENT}Crystallographic data and refinement parameters for \emph{HE}166 at room temperature.} \begin{tabular}{cc} \hline \makebox[0.45\textwidth][c]{Chemical Formula} & \makebox[0.45\textwidth][c]{\ch{Tb_{0.18}Dy_{0.15}Ho_{0.17}Er_{0.19}Tm_{0.17}Lu_{0.12}Mn_{6}Sn_{6}}} \\ \makebox[0.45\textwidth][c]{Space Group} & \makebox[0.45\textwidth][c]{$P6/mmm$} \\ \makebox[0.45\textwidth][c]{\textit{a} (\AA)} & \makebox[0.45\textwidth][c]{5.5173 (6)} \\ \makebox[0.45\textwidth][c]{\textit{c} (\AA)} & \makebox[0.45\textwidth][c]{9.0053 (11)} \\ \makebox[0.45\textwidth][c]{\textit{V} (\AA\textsuperscript{3})} & \makebox[0.45\textwidth][c]{237.40 (6)} \\ \makebox[0.45\textwidth][c]{\textit{Z}} & \makebox[0.45\textwidth][c]{1} \\ \makebox[0.45\textwidth][c]{Density (g/cm\textsuperscript{3})} & \makebox[0.45\textwidth][c]{8.424} \\ \makebox[0.45\textwidth][c]{Abs. Coefficient (mm\textsuperscript{-1})} & \makebox[0.45\textwidth][c]{31.41} \\ \makebox[0.45\textwidth][c]{Crystal Size (mm\textsuperscript{3})} & \makebox[0.45\textwidth][c]{0.06$\times$0.03$\times$0.01} \\ \makebox[0.45\textwidth][c]{$\theta$ Range ($^\circ$)} & \makebox[0.45\textwidth][c]{2.3-30.3} \\ \makebox[0.45\textwidth][c]{Index Range} & \makebox[0.45\textwidth][c]{} \\ \makebox[0.45\textwidth][c]{$h$} & \makebox[0.45\textwidth][c]{-7 $\to$ 7} \\ \makebox[0.45\textwidth][c]{$k$} & \makebox[0.45\textwidth][c]{-7 $\to$ 7} \\ \makebox[0.45\textwidth][c]{$l$} & \makebox[0.45\textwidth][c]{-12 $\to$ 12} \\ \makebox[0.45\textwidth][c]{Number of Reflections} & \makebox[0.45\textwidth][c]{9986} \\ \makebox[0.45\textwidth][c]{Unique Reflections} & \makebox[0.45\textwidth][c]{187} \\ \makebox[0.45\textwidth][c]{Data / Restraints / Parameters} & \makebox[0.45\textwidth][c]{187 / 0 / 15} \\ \makebox[0.45\textwidth][c]{$R_{\text{int}}$} & \makebox[0.45\textwidth][c]{0.051} \\ \makebox[0.45\textwidth][c]{$\Delta\rho_{\text{max}}$ / $\Delta\rho_{\text{min}}$ (e\AA \textsuperscript{-3})} & \makebox[0.45\textwidth][c]{1.74 / -1.36}\\ \makebox[0.45\textwidth][c]{GoF} & \makebox[0.45\textwidth][c]{1.24} \\ \makebox[0.45\textwidth][c]{$R_1[F^2>2\sigma(F^2)]$} & \makebox[0.45\textwidth][c]{0.017} \\ \makebox[0.45\textwidth][c]{$wR(F^2)$} & \makebox[0.45\textwidth][c]{0.046} \\ \hline \end{tabular} \end{table*} 

\begin{table*}[ht] \centering \caption{\label{tab:ATOMIC_COORDINATES} Fractional atomic coordinates and isotropic or equivalent isotropic displacement parameters.} \begin{tabular}{cccccc} \hline\\[-2ex] \makebox[0.16\textwidth][c]{Atom} & \makebox[0.16\textwidth][c]{x} & \makebox[0.16\textwidth][c]{y} & \makebox[0.16\textwidth][c]{z} & \makebox[0.16\textwidth][c]{$U_{\text{iso}}*/U_{\text{eq}}$ (\AA$^2$)} & \makebox[0.16\textwidth][c]{Occ. ($<1$)} \\ \hline \makebox[0.16\textwidth][c]{Tb1} & \makebox[0.16\textwidth][c]{0} & \makebox[0.16\textwidth][c]{0} & \makebox[0.16\textwidth][c]{0} & \makebox[0.16\textwidth][c]{0.0088(2)} & \makebox[0.16\textwidth][c]{0.18} \\ \makebox[0.16\textwidth][c]{Dy1} & \makebox[0.16\textwidth][c]{0} & \makebox[0.16\textwidth][c]{0} & \makebox[0.16\textwidth][c]{0} & \makebox[0.16\textwidth][c]{0.0088(2)} & \makebox[0.16\textwidth][c]{0.15} \\ \makebox[0.16\textwidth][c]{Ho1} & \makebox[0.16\textwidth][c]{0} & \makebox[0.16\textwidth][c]{0} & \makebox[0.16\textwidth][c]{0} & \makebox[0.16\textwidth][c]{0.0088(2)} & \makebox[0.16\textwidth][c]{0.17} \\ \makebox[0.16\textwidth][c]{Er1} & \makebox[0.16\textwidth][c]{0} & \makebox[0.16\textwidth][c]{0} & \makebox[0.16\textwidth][c]{0} & \makebox[0.16\textwidth][c]{0.0088(2)} & \makebox[0.16\textwidth][c]{0.19} \\ \makebox[0.16\textwidth][c]{Tm1} & \makebox[0.16\textwidth][c]{0} & \makebox[0.16\textwidth][c]{0} & \makebox[0.16\textwidth][c]{0} & \makebox[0.16\textwidth][c]{0.0088(2)} & \makebox[0.16\textwidth][c]{0.17} \\ \makebox[0.16\textwidth][c]{Lu1} & \makebox[0.16\textwidth][c]{0} & \makebox[0.16\textwidth][c]{0} & \makebox[0.16\textwidth][c]{0} & \makebox[0.16\textwidth][c]{0.0088(2)} & \makebox[0.16\textwidth][c]{0.12} \\ \makebox[0.16\textwidth][c]{Mn1} & \makebox[0.16\textwidth][c]{$\frac{1}{2}$} & \makebox[0.16\textwidth][c]{$\frac{1}{2}$} & \makebox[0.16\textwidth][c]{0.24727(8)} & \makebox[0.16\textwidth][c]{0.0067(2)} & \makebox[0.16\textwidth][c]{} \\ \makebox[0.16\textwidth][c]{Sn1} & \makebox[0.16\textwidth][c]{0} & \makebox[0.16\textwidth][c]{0} & \makebox[0.16\textwidth][c]{0.33660(6)} & \makebox[0.16\textwidth][c]{0.0089(2)} & \makebox[0.16\textwidth][c]{} \\ \makebox[0.16\textwidth][c]{Sn2} & \makebox[0.16\textwidth][c]{$\frac{2}{3}$} & \makebox[0.16\textwidth][c]{$\frac{1}{3}$} & \makebox[0.16\textwidth][c]{0} & \makebox[0.16\textwidth][c]{0.0077(2)} & \makebox[0.16\textwidth][c]{} \\[0.5ex] \makebox[0.16\textwidth][c]{Sn3} & \makebox[0.16\textwidth][c]{$\frac{2}{3}$} & \makebox[0.16\textwidth][c]{$\frac{1}{3}$} & \makebox[0.16\textwidth][c]{$\frac{1}{2}$} & \makebox[0.16\textwidth][c]{0.0088(2)} & \makebox[0.16\textwidth][c]{} \\[0.5ex] \hline\\ \end{tabular} \end{table*} 

\begin{table*}[ht] \centering \caption{\label{tab:ATOMIC_DISPLACEMENT} Atomic displacement parameters (\AA$^2$).} \begin{tabular}{ccccccc} \hline\\[-2ex] \makebox[0.135\textwidth][c]{Atom} & \makebox[0.135\textwidth][c]{$U^{11}$} & \makebox[0.135\textwidth][c]{$U^{22}$} & \makebox[0.135\textwidth][c]{$U^{33}$} & \makebox[0.135\textwidth][c]{$U^{12}$} & \makebox[0.135\textwidth][c]{$U^{13}$} & \makebox[0.135\textwidth][c]{$U^{23}$} \\ \hline \makebox[0.135\textwidth][c]{Tb1} & \makebox[0.135\textwidth][c]{0.0084(2)} & \makebox[0.135\textwidth][c]{0.0084(2)} & \makebox[0.135\textwidth][c]{0.0096(3)} & \makebox[0.135\textwidth][c]{0.00419(12)} & \makebox[0.135\textwidth][c]{0} & \makebox[0.135\textwidth][c]{0} \\ \makebox[0.135\textwidth][c]{Dy1} & \makebox[0.135\textwidth][c]{0.0084(2)} & \makebox[0.135\textwidth][c]{0.0084(2)} & \makebox[0.135\textwidth][c]{0.0096(3)} & \makebox[0.135\textwidth][c]{0.00419(12)} & \makebox[0.135\textwidth][c]{0} & \makebox[0.135\textwidth][c]{0} \\ \makebox[0.135\textwidth][c]{Ho1} & \makebox[0.135\textwidth][c]{0.0084(2)} & \makebox[0.135\textwidth][c]{0.0084(2)} & \makebox[0.135\textwidth][c]{0.0096(3)} & \makebox[0.135\textwidth][c]{0.00419(12)} & \makebox[0.135\textwidth][c]{0} & \makebox[0.135\textwidth][c]{0} \\ \makebox[0.135\textwidth][c]{Er1} & \makebox[0.135\textwidth][c]{0.0084(2)} & \makebox[0.135\textwidth][c]{0.0084(2)} & \makebox[0.135\textwidth][c]{0.0096(3)} & \makebox[0.135\textwidth][c]{0.00419(12)} & \makebox[0.135\textwidth][c]{0} & \makebox[0.135\textwidth][c]{0} \\ \makebox[0.135\textwidth][c]{Tm1} & \makebox[0.135\textwidth][c]{0.0084(2)} & \makebox[0.135\textwidth][c]{0.0084(2)} & \makebox[0.135\textwidth][c]{0.0096(3)} & \makebox[0.135\textwidth][c]{0.00419(12)} & \makebox[0.135\textwidth][c]{0} & \makebox[0.135\textwidth][c]{0} \\ \makebox[0.135\textwidth][c]{Lu1} & \makebox[0.135\textwidth][c]{0.0084(2)} & \makebox[0.135\textwidth][c]{0.0084(2)} & \makebox[0.135\textwidth][c]{0.0096(3)} & \makebox[0.135\textwidth][c]{0.00419(12)} & \makebox[0.135\textwidth][c]{0} & \makebox[0.135\textwidth][c]{0} \\ \makebox[0.135\textwidth][c]{Mn1} & \makebox[0.135\textwidth][c]{0.0073(3)} & \makebox[0.135\textwidth][c]{0.0073(3)} & \makebox[0.135\textwidth][c]{0.0064(4)} & \makebox[0.135\textwidth][c]{0.0043(3)} & \makebox[0.135\textwidth][c]{0} & \makebox[0.135\textwidth][c]{0} \\ \makebox[0.135\textwidth][c]{Sn1} & \makebox[0.135\textwidth][c]{0.0089(3)} & \makebox[0.135\textwidth][c]{0.0089(3)} & \makebox[0.135\textwidth][c]{0.0088(3)} & \makebox[0.135\textwidth][c]{0.00447(13)} & \makebox[0.135\textwidth][c]{0} & \makebox[0.135\textwidth][c]{0} \\ \makebox[0.135\textwidth][c]{Sn2} & \makebox[0.135\textwidth][c]{0.0083(3)} & \makebox[0.135\textwidth][c]{0.0083(3)} & \makebox[0.135\textwidth][c]{0.0066(3)} & \makebox[0.135\textwidth][c]{0.00414(13)} & \makebox[0.135\textwidth][c]{0} & \makebox[0.135\textwidth][c]{0} \\ \makebox[0.135\textwidth][c]{Sn3} & \makebox[0.135\textwidth][c]{0.0100(3)} & \makebox[0.135\textwidth][c]{0.0100(3)} & \makebox[0.135\textwidth][c]{0.0065(3)} & \makebox[0.135\textwidth][c]{0.00498(13)} & \makebox[0.135\textwidth][c]{0} & \makebox[0.135\textwidth][c]{0} \\ \hline\\ \end{tabular} \end{table*} 

\begin{table*}[ht] \centering \caption{\label{tab:INTERATOMIC_DISTANCES} Relevant interatomic distances.} \begin{tabular}{cc} \hline\\[-2ex] \makebox[0.45\textwidth][c]{Atoms} & \makebox[0.45\textwidth][c]{Distance (\AA)} \\ \hline \makebox[0.45\textwidth][c]{Mn1-Mn1 (in-plane)} & \makebox[0.45\textwidth][c]{2.7587(3)} \\ \makebox[0.45\textwidth][c]{Mn1-Mn1 ($J_1$)} & \makebox[0.45\textwidth][c]{4.5518(15)} \\ \makebox[0.45\textwidth][c]{Mn1-Mn1 ($J_2$)} & \makebox[0.45\textwidth][c]{4.4535(15)} \\ \makebox[0.45\textwidth][c]{Mn1-Mn1 ($J_3$)} & \makebox[0.45\textwidth][c]{9.0053(11)} \\ \makebox[0.45\textwidth][c]{Mn1-Sn1} & \makebox[0.45\textwidth][c]{2.8736(4)} \\ \makebox[0.45\textwidth][c]{Mn1-Sn2} & \makebox[0.45\textwidth][c]{2.7377(6)} \\ \makebox[0.45\textwidth][c]{Mn1-Sn3} & \makebox[0.45\textwidth][c]{2.7778(6)} \\ \makebox[0.45\textwidth][c]{Sn1-Sn1} & \makebox[0.45\textwidth][c]{2.9429(11)} \\ \makebox[0.45\textwidth][c]{Sn2-Sn2} & \makebox[0.45\textwidth][c]{3.1854(4)} \\ \makebox[0.45\textwidth][c]{Sn3-Sn3} & \makebox[0.45\textwidth][c]{3.1854(4)} \\ \makebox[0.45\textwidth][c]{\emph{R}1-Sn1} & \makebox[0.45\textwidth][c]{3.0312(7)} \\ \makebox[0.45\textwidth][c]{\emph{R}1-Sn2} & \makebox[0.45\textwidth][c]{3.1854(4)} \\ \makebox[0.45\textwidth][c]{\emph{R}1-Mn1} & \makebox[0.45\textwidth][c]{3.5452(5)} \\ \makebox[0.45\textwidth][c]{\emph{R}1-Mn plane ($J_{R-\textrm{Mn}}$)} & \makebox[0.45\textwidth][c]{2.2267(7)} \\ \hline\\ \end{tabular} \end{table*} 

\subsection{EDX}

\begin{figure*}
\centering
  \includegraphics[width=.8
  \textwidth]{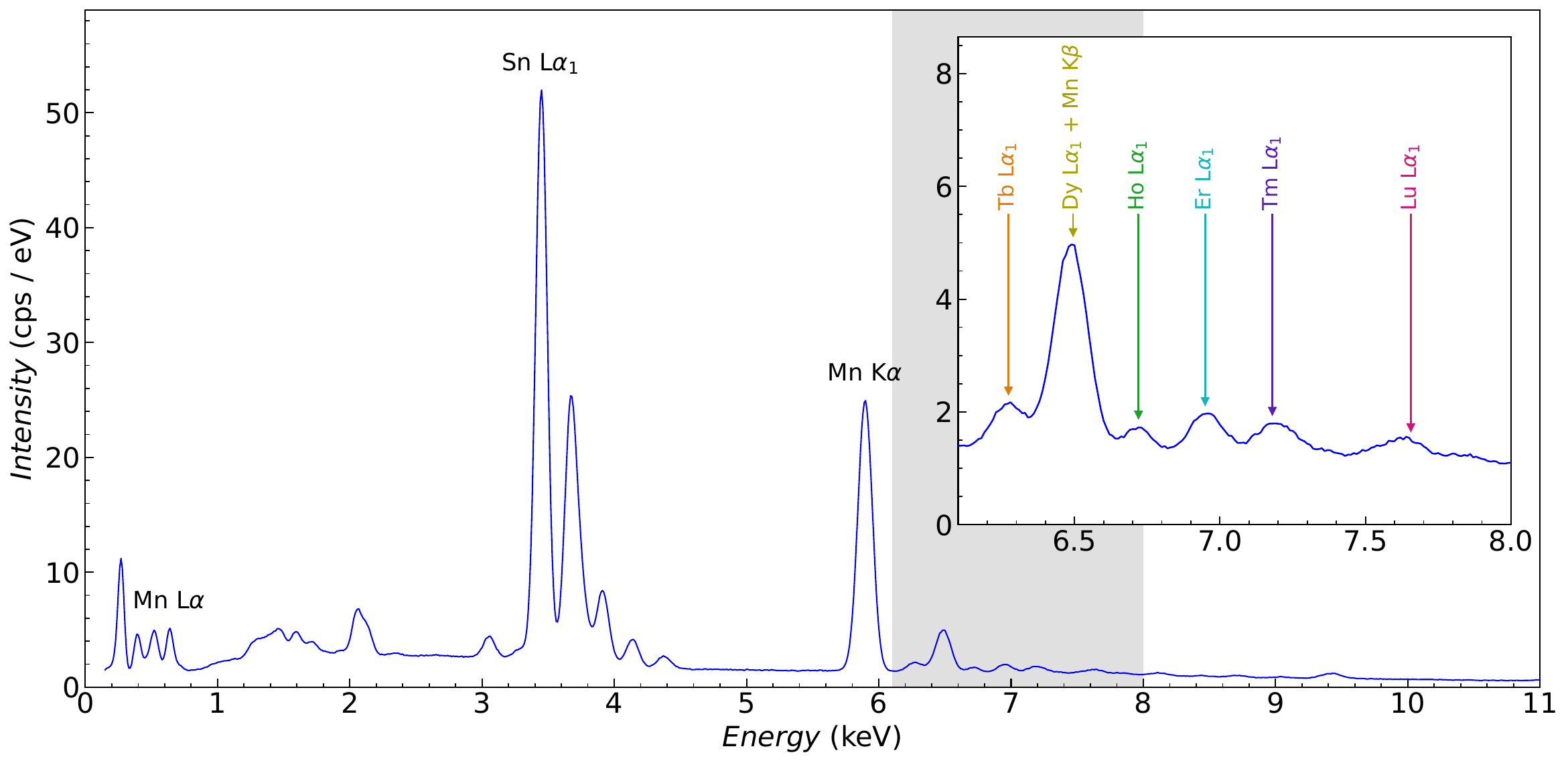}
  \caption{\label{fig:EDX_SPECTRUM}
    Example EDX spectrum of \emph{HE}166. Inset: zoomed-in view highlighting the separate rare-earth peaks.
  }
\end{figure*}

EDX was used to confirm the sample stoichiometry, which was determined to be {\footnotesize \ch{Tb_{0.18(2)}Dy_{0.15(1)}Ho_{0.17(1)}Er_{0.19(1)}Tm_{0.17(1)}Lu_{0.12(2)}Mn_{6.2(2)}Sn_{6.1(2)}}}.
The reported error represents the standard deviation over scans of multiple different sites.
An example EDX spectrum highlighting the separated L$\alpha_1$ peaks of the various rare-earths is shown in Fig.~\ref{fig:EDX_SPECTRUM}.

\begin{figure*}
\centering
  \includegraphics[width=\textwidth]{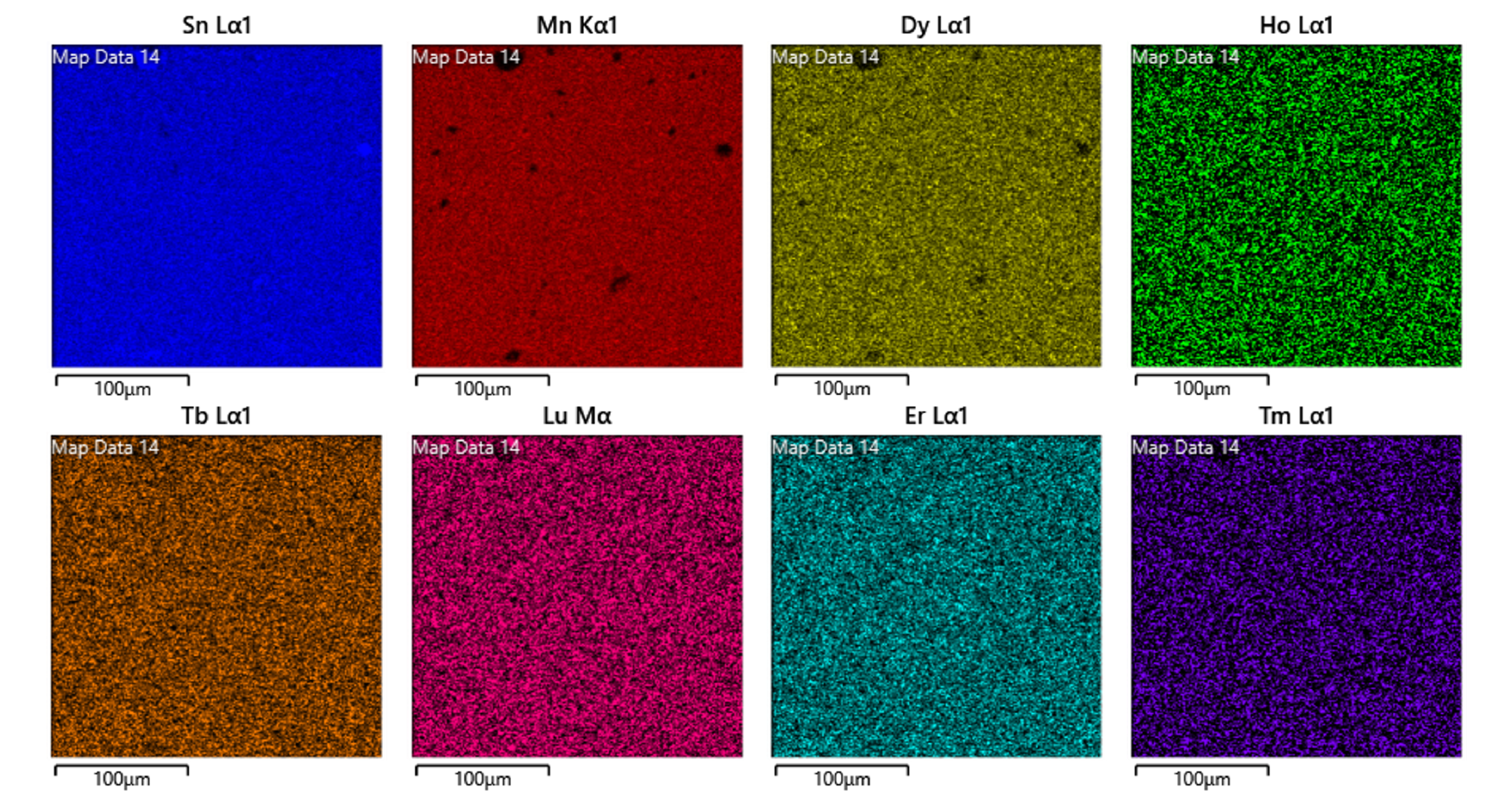}
  \caption{\label{fig:250NM_COLORMAP}
    EDX colormap showing distribution of elements making up \emph{HE}166. The colormap is 255$\times$250 $\mu$m.
  }
\end{figure*}

\begin{figure*}
\centering
  \includegraphics[width=\textwidth]{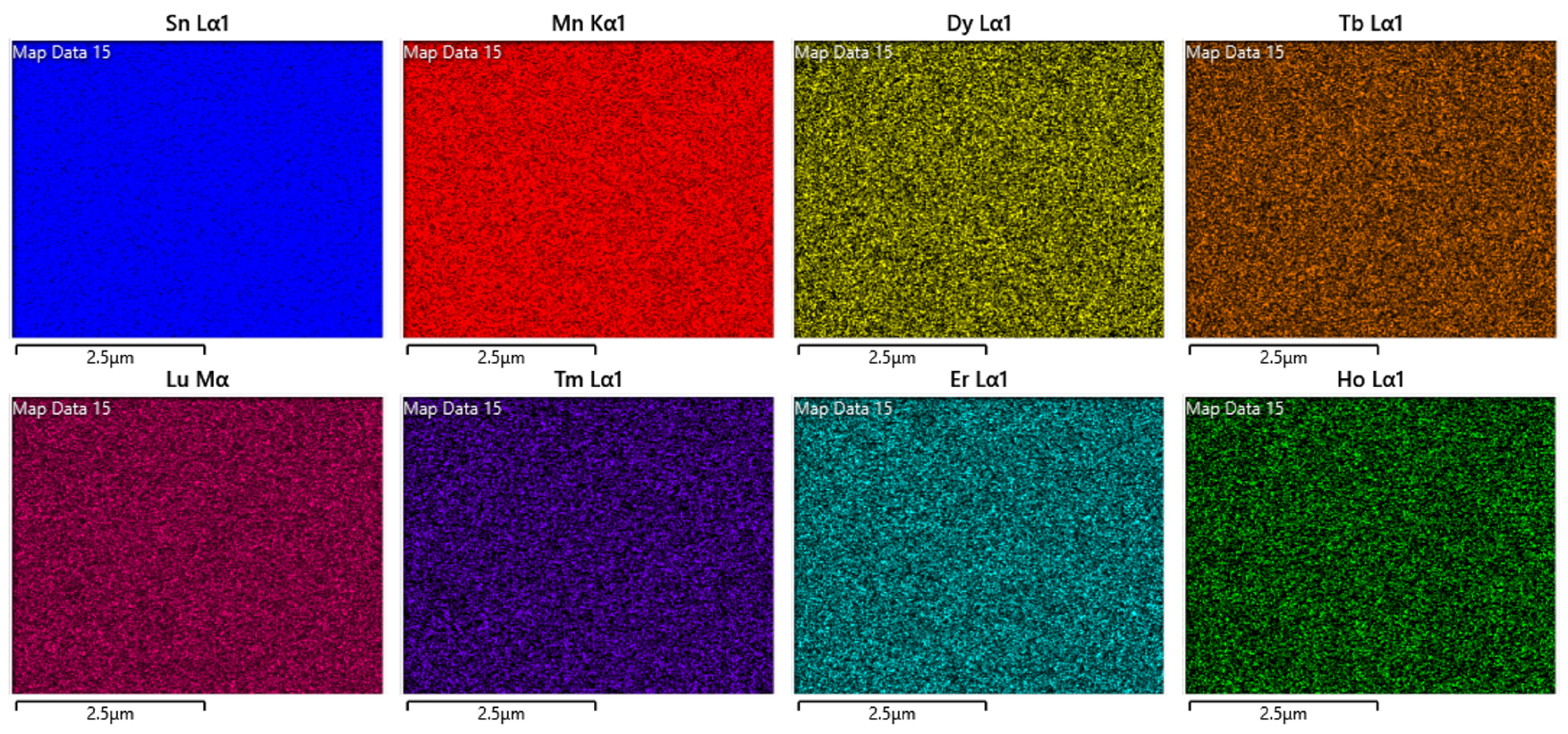}
  \caption{\label{fig:5NM_COLORMAP}
    EDX colormaps showing the distribution of elements making up \emph{HE}166. The colormap is 5$\times$4 $\mu$m.
  }
\end{figure*}

EDX colormaps in Figs.~\ref{fig:250NM_COLORMAP} and \ref{fig:5NM_COLORMAP} reveal that all rare earth elements are present without phase separation on either a 250 $\mu$m scale or 5 $\mu$m scale.

\section{Magnetization Characterization}

\subsection{Transition Temperature Determination}

\begin{figure*}
\centering
  \includegraphics[width=.6\textwidth]{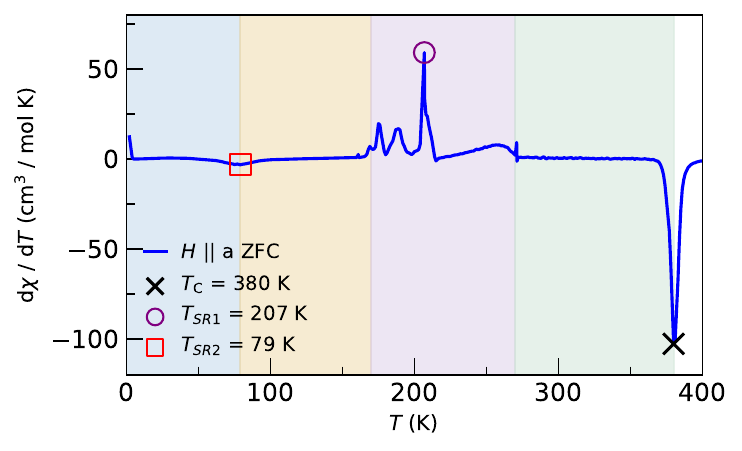}
  \caption{\label{fig:SUS_DERIVATIVE}
    Derivative of the DC susceptibility of \emph{HE}166 (Fig.~1d of main text) for the ZFC data with the field applied parallel to the $a$-axis. Transition temperatures determined by minima and maxima in transition regions. $T_{\text{SR1}}$ was determined from the highest peak in the transitionary region.
  }
\end{figure*}

The transition temperatures \TC, $T_{\text{SR1}}$, and $T_{\text{SR2}}$ were determined from the maxima and minima in the derivative of the ZFC $H||a$ susceptibility data as shown in Fig.~\ref{fig:SUS_DERIVATIVE}. 
The error in the reported values is the full width at half maximum (or minimum) of these extrema.

\subsection{Magnetic Phase Diagrams}

\begin{figure*}
\centering
  \includegraphics[width=.89\textwidth]{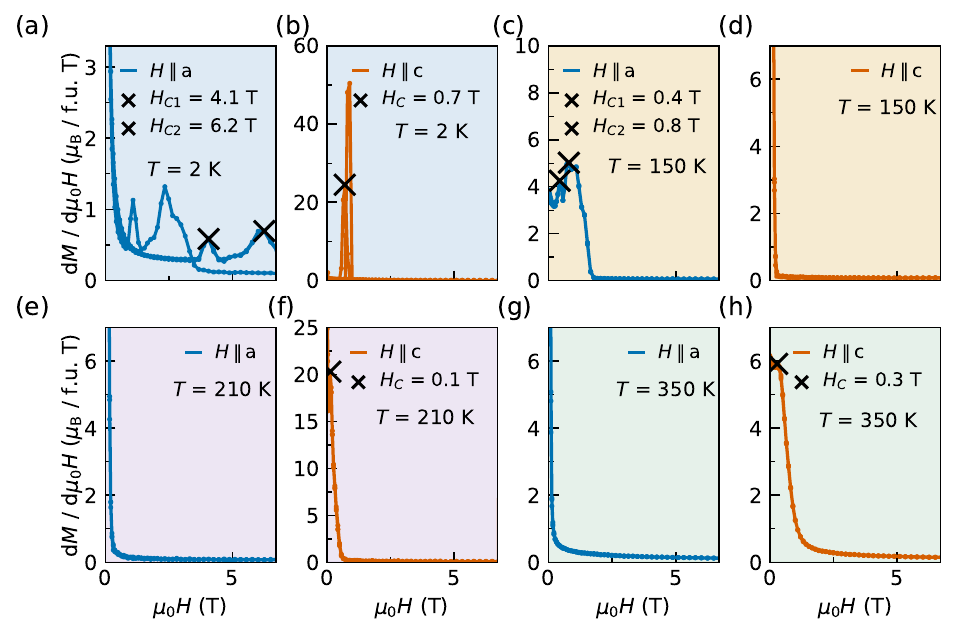}
  \caption{\label{fig:dM(H)dH}
    Derivative of the magnetization of $HE$166 with respect to field at (a,b) 2 K (c,d) 150 K (e,f) 210 K and (g,h) 350 K.
  }
\end{figure*}

The field induced transitions between states were determined using peaks in the derivative of the magnetization with respect to field as shown in Fig.~\ref{fig:dM(H)dH}.
When there is significant hysteresis in the critical field, the value measured in a zero-field cooled increasing field set-up was used.
In the easy-plane region ($T \gtrsim$ 270 K, green) there is not a true metamagnetic transition when the field is applied along the c-axis; however, a border is still placed at the field corresponding to a sharp decrease in the derivative where the magnetization begins to saturate.
This follows the trend in the lower temperature regions as the metamagnetic transition moves closer to the saturation field as the temperature increases.

\begin{figure*}
\centering
  \includegraphics[width=.89\textwidth]{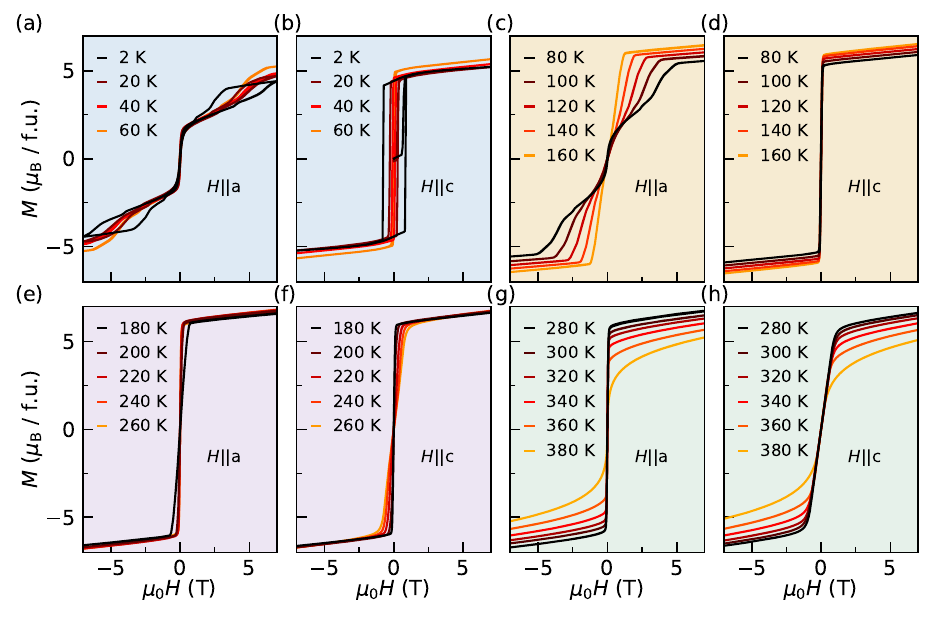}
  \caption{\label{fig:MAG_EVO}
    Evolution of the magnetization as a function of field for $HE$166 in (a,b) the canted low-temperature region, (c,d) the easy-axis region, (e,f) the transitional region, and (g,h) the easy-plane region.
  }
\end{figure*}

Fig.~\ref{fig:MAG_EVO} shows the evolution in the magnetization behavior of $HE$166.
As discussed in the main text, in the low temperature canted region (blue) there is significant hysteresis with field-induced transitions regardless of the applied field direction.
In the easy-axis (orange) region $HE$166 is a magnetically softer material with negligible hysteresis.
There are still two metamagnetic transitions when $H||a$, however there are no longer any field induced transitions when $H||c$.
In the transitional region (purple) there is a metamagnetic transition at low fields when the field is applied along the $c$-axis but no transition can now be detected when the field is applied in-plane.
In the easy-plane region (green), there do not appear any transitions for either applied field direction.

From the critical fields determined, two phase diagrams were produced as shown in the main text Fig.~3. 
Above 270 K, there is no longer a clear peak in the magnetization curve correlating to a metamagnetic transition with field applied along the $c$-axis, therefore, the border between the in-plane and axis aligned spins are taken at the point of spin saturation, determined by the beginning of a sharp decrease in the derivative, this follows the general trend as the transition peak that is apparent at lower temperatures occurs just before this drop off and remains there as its prominence decreases until it is undetectable.

\section{Electrical Transport Characterization}

\subsection{Longitudinal Resistivity}

\begin{figure*}
\centering
  \includegraphics[width=\textwidth]{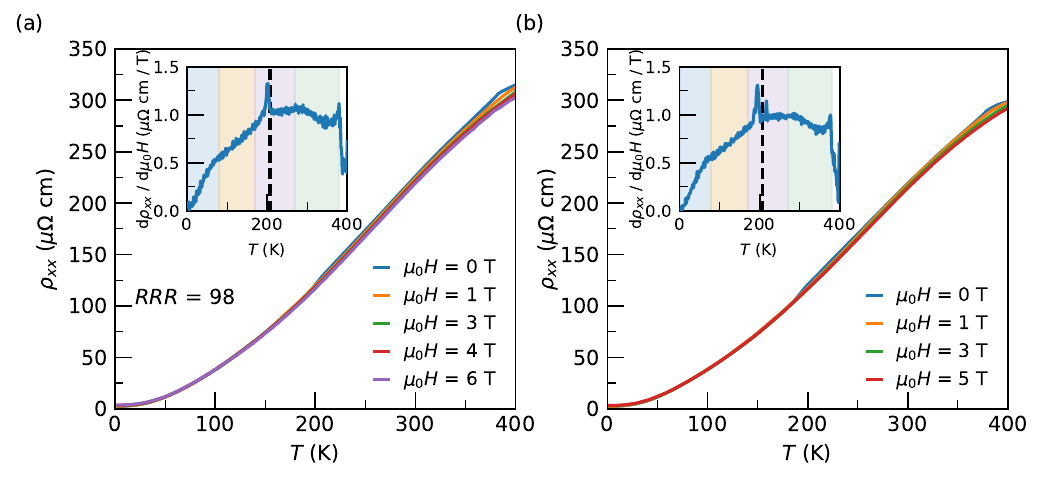}
  \caption{\label{fig:R(T)}
    Resistivity as a function of temperature as measured at multiple fields for the magnetic field (a) applied along the $a$-axis and (b) applied along the $c$-axis. Insets show the derivative of resistivity with respect to temperature for the zero-field measurements.
    The black dashed lines mark $T_{\text{SR1}}$.
  }
\end{figure*}

In order to determine the quality of the crystal and to confirm the determined transition temperatures the longitudinal resistivity was measured as a function of temperature at multiple fields as shown in Fig.~\ref{fig:R(T)}.
The residual resistivity ratio (RRR) as determined by the zero-field resistivity at 300 K and 2 K had a value of 98, characteristic of a fairly high-quality crystal.
The insets of Fig.~\ref{fig:R(T)} show the derivative of the zero-field measured resistivities showing a consistent peak near 380 K and at 200 K. 
These agree well with the \TC~and $T_{\text{SR1}}$ values determined from magnetization data. 
Unfortunately, the low temperature transition is not clear from this first derivative.

\subsection{Hall Effect}

\begin{figure*}
\centering
  \includegraphics[width=.8\textwidth]{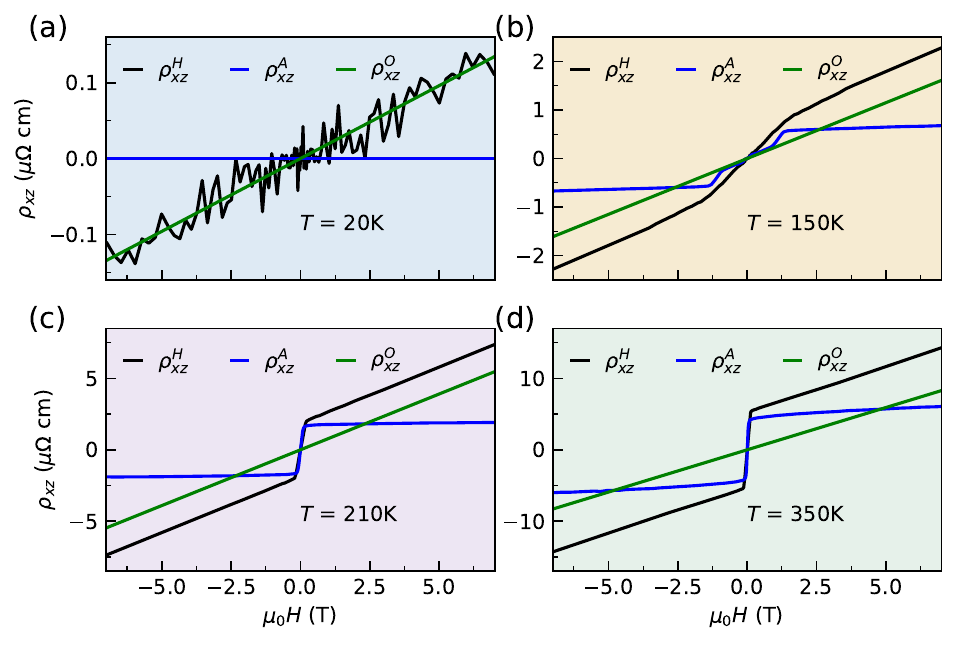}
  \caption{\label{fig:HALL_HA}
    Decomposed Hall measurements for $HE$166 with the field applied along the $a$-axis at (a) 20 K (b) 150 K (c) 210 K and (d) 350 K.
  }
\end{figure*}

\begin{figure*}
\centering
  \includegraphics[width=.8\textwidth]{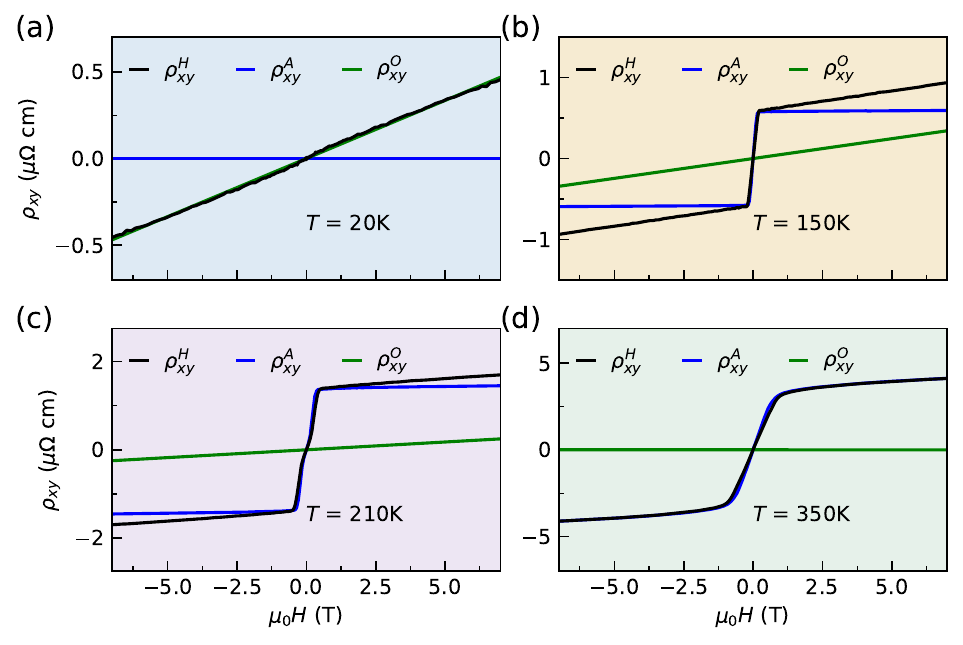}
  \caption{\label{fig:HALL_HC}
    Decomposed Hall measurements for $HE$166 with the field applied along the $c$-axis at (a) 20 K (b) 150 K (c) 210 K and (d) 350 K.
  }
\end{figure*}

The Hall effect was studied using a four-contact transverse wiring geometry. The transverse resistivity data was first anti-symmetrized in order to remove and longitudinal component. Then, at high-fields, the ordinary Hall coefficient and anomalous Hall coefficient were extracted by assuming that $\rho_H=\rho_H^O+\rho_H^A=R_0H+4\pi R_SM$. Using these coefficients the Hall resistivity was then decomposed into an ordinary Hall resistivity and anomalous Hall resistivity. Figs.~\ref{fig:HALL_HA} \& \ref{fig:HALL_HC} show this decomposition for field applied along both the a-axis and the c-axis. The Hall effect behaves comparably to previous reports and does not seem to have been changed by the inclusion of spiral rare-earths~[3].

\begin{figure*}
\centering
  \includegraphics[width=.6\textwidth]{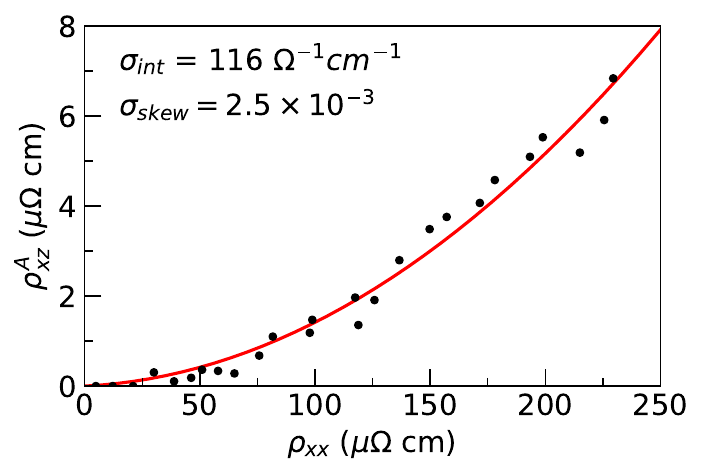}
  \caption{\label{fig:TYJ}
    TYJ scaling relation for $HE$166 as determined from Hall effect decomposition when field is applied along the $a$-axis.
  }
\end{figure*}

From this decomposition it is possible to do Tian-Ye-Jin (TYJ) analysis in order to determine if the anomalous Hall effect is primarily intrinsic or due to impurity caused skew scattering. 
Fig.~\ref{fig:TYJ} shows this analysis where the anomalous Hall effect is plotted against the longitudinal resistivity and then fitted to the TYJ scaling law: $\rho^A_{zx}=\sigma_{int}\rho^2_{xx}+\sigma_{skew}\rho_{xx}$.
This measured data is dominated by the parabolic term pointing to a dominance of the intrinsic anomalous Hall effect.
From the $\sigma_{int}$ extracted from this fit, the conductance per Mn layer can be calculated from the intrinsic conductivity to be 0.14$\frac{e^2}{\hbar}$.
This value is comparable to other \ch{$R$Mn6Sn6} materials and previous high-entropy reports, so the mixing of different rare-earths does not result in a fundamentally different behavior in the intrinsic conductance.

\section{Neutron Diffraction}

\subsection{Single-crystal refinement}

Integrated single-crystal neutron intensities were obtained from ZEBRA $\omega$ scans at 300~K, 150~K, and 6~K. The peak intensities were Lorentz corrected and then corrected for absorption using the measured plate-like crystal dimensions, the UB orientation matrix, and the reflection-dependent incident and exit beam paths through the crystal. The crystal was treated as an approximately $4\times2.5\times0.9$~mm$^3$ plate with the $c$ axis normal to the plate. The 300~K and 150~K refinements used the room-temperature single-crystal x-ray structure, while the 6~K refinement used the same atomic positions with atomic displacement parameters set to zero.

The nuclear scale factor was refined from the nuclear reflections and held fixed for the magnetic refinements. 
Magnetic and nuclear structure factors were calculated using the FullProf/CrysFML magnetic structure-factor machinery through Python bindings rather than by refining PCR files~[24, 25]. 
A plate-like anisotropic extinction correction was applied using the CrysFML/FullProf implementation of the SHELX extinction form~[21] with
\begin{equation}
{\rm ext} = \left[{\rm ext}_{ab},{\rm ext}_{ab},0,0,0,0\right],
\end{equation}
where the six entries are the anisotropic extinction tensor coefficients in the SHELX convention, here constrained to refine only a common basal-plane coefficient.

The magnetic refinement used a $q=0$ model with a coherent in-plane Mn order parameter spanned by basis vectors of the 191.238 and 191.239 magnetic space groups~[27, 28],
\begin{equation}
{\bf M}^{\rm Mn}_{ab}=M^{\rm Mn}_{ab}\left[\cos(\alpha)\psi_{238}+\sin(\alpha)\psi_{239}\right].
\end{equation}
The basis vectors $\psi_{238}$ and $\psi_{239}$ are orthogonal components of the same 120$^\circ$ in-plane Mn order parameter. The model also included a uniform Mn $c$-axis component and a canted effective rare-earth moment with independent in-plane and $c$-axis components. For the effective rare-earth form factor, the Ho$^{3+}$ rare-earth form factor was used as a representative average for the mixed rare-earth site. Because the measurement is unpolarized and the rare-earth site is chemically disordered, the refined rare-earth moment should be interpreted as an effective site-averaged moment.

\begin{figure*}[t]
\centering
\includegraphics[width=\textwidth]{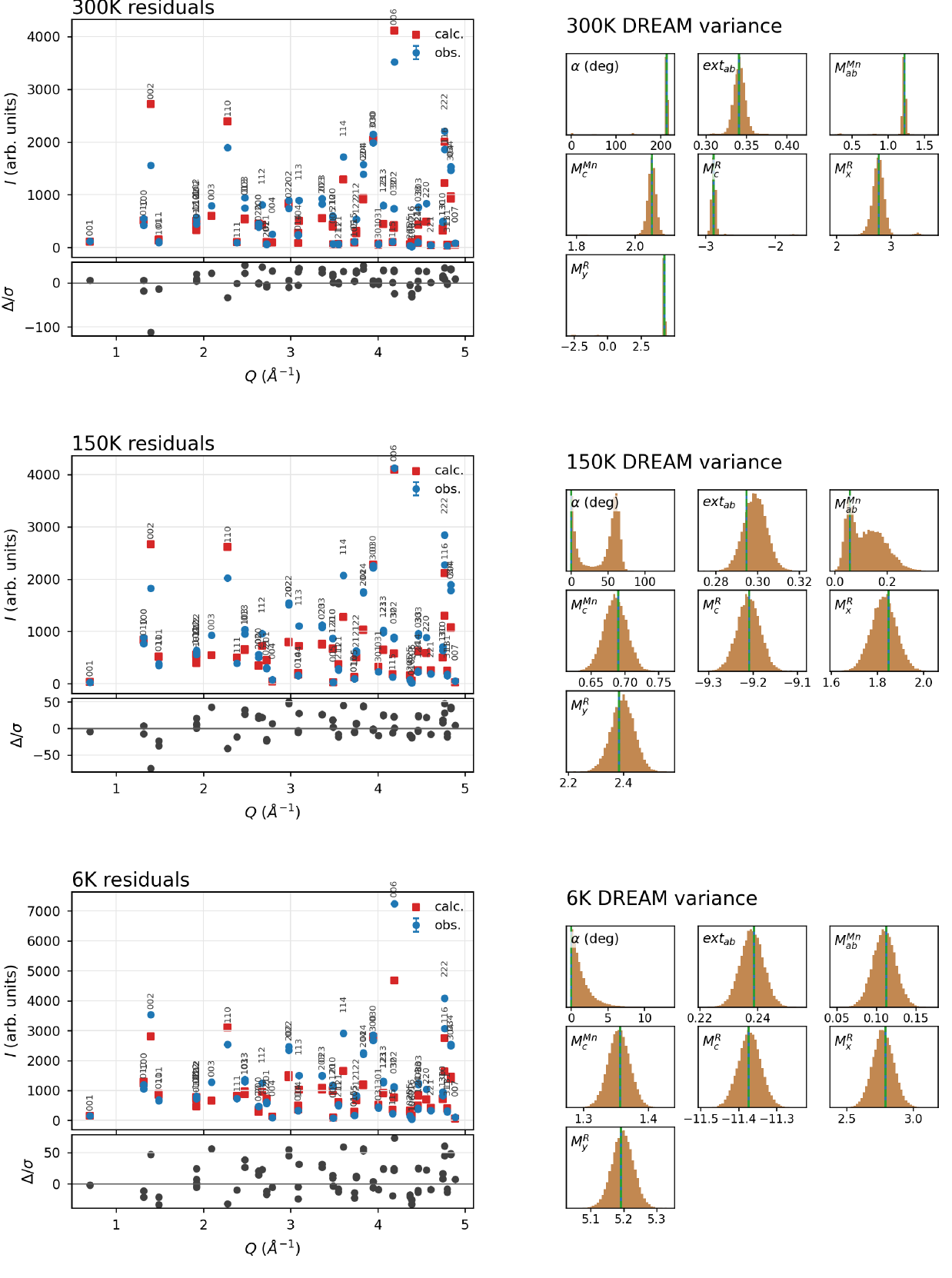}
\caption{Coherent $U(1)$ magnetic refinement diagnostics for 300~K, 150~K, and 6~K. For each temperature, the left panel shows observed and calculated $F^2$ values as a function of $Q$ with residuals below, and the right panel shows the corresponding Bumps/DREAM variance plot.}
\label{fig:U1_REFINEMENT_DIAGNOSTICS}
\end{figure*}

The outer refinement loop used Bumps with the DREAM algorithm~[26]. DREAM is a population Markov-chain sampler designed for global exploration of multimodal parameter spaces, which is useful here because magnetic refinements can have symmetry-related minima and strong parameter correlations. The Bumps variance plots in Fig.~\ref{fig:U1_REFINEMENT_DIAGNOSTICS} show the marginal posterior distributions after burn-in together with the best-likelihood profile for each parameter; the green line marks the best-fit value used for the calculated intensities.

\begin{table*}[t]
\setlength{\tabcolsep}{4.5pt}
\begin{tabular}{cccccccccc}
\hline\hline
Temp. & $\alpha$ & $M^{\rm Mn}_{ab}$ & $M^{\rm Mn}_c$ & $M^R_{ab}$ & $M^R_c$ & $\theta^R$ & ${\rm ext}_{ab}$ & $R_{\rm abs}$ & WRMS \\
(K) & ($^\circ$) & ($\mu_B$) & ($\mu_B$) & ($\mu_B$) & ($\mu_B$) & ($^\circ$) & & & \\
\hline
300 & 213.2 & 1.22 & 2.06 & 5.04 & -2.88 & 60.2 & 0.341 & 0.286 & 23.59 \\
150 & 0.09 & 0.06 & 0.69 & 3.02 & -9.21 & 18.1 & 0.295 & 0.285 & 24.23 \\
6   & 0.005 & 0.11 & 1.36 & 5.89 & -11.37 & 27.4 & 0.239 & 0.294 & 28.14 \\
\hline\hline
\end{tabular}
\caption{Parameters from the coherent $U(1)$ basis-vector refinement. The angle $\alpha$ defines the in-plane Mn basis-vector phase. The angle $\theta^R$ is the canting angle of the effective rare-earth moment away from the dominant $c$-axis direction. WRMS is the root-mean-square residual in units of the experimental uncertainty.}
\label{tab:U1_REFINEMENT_PARAMETERS}
\end{table*}

As a model comparison, we also tested an incoherent two-phase model built from the same normalized 191.238 and 191.239 endpoint configurations. 
The coherent $U(1)$ model and the incoherent population model have the same number of fitted parameters. 
At 300~K, the coherent $U(1)$ model gives a slightly lower residual ($R_{\rm abs}$),  $R_{\rm abs}=0.286$, WRMS=23.59, than the incoherent population model, $R_{\rm abs}=0.296$, WRMS=23.84, while at 150~K and 6~K the two models are statistically indistinguishable because the fitted in-plane Mn amplitude is very small and the $U(1)$ angle has little leverage. 
Thus the $U(1)$ description is the preferred compact basis-vector language for the main text.

\subsection{Small-Angle Neutron diffraction}

\begin{figure*}[t]
\centering
\includegraphics[width=\textwidth]{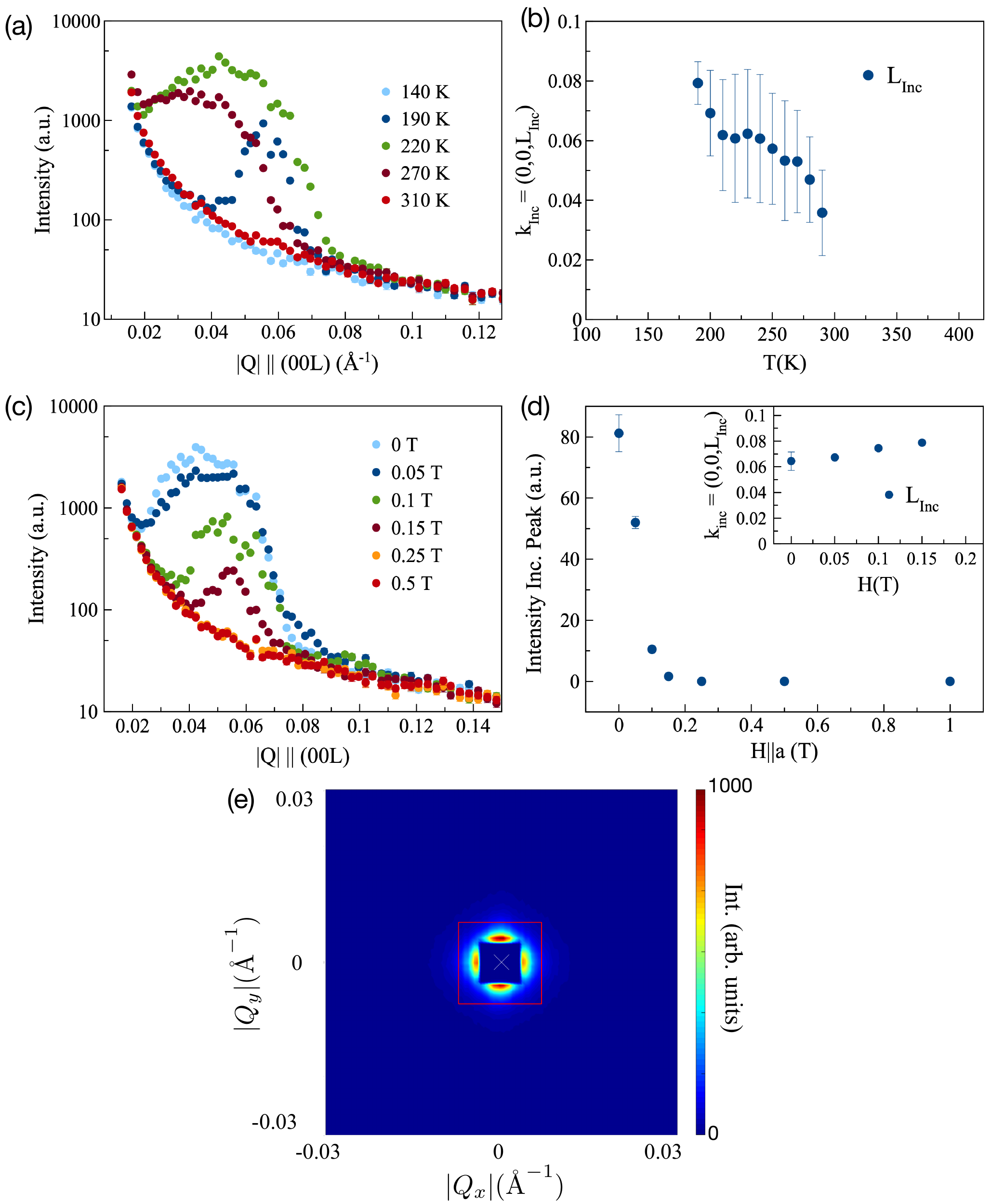}
\caption{(a) 1D cuts showing the intensity as function of momentum transfer parallel to the $(00L)$ reciprocal space direction collected in zero-field and for several temperatures. (b) Zero-field temperature dependence of the wave vector incommensurate modulation obtained from fitting position of the incommensurate peak seen in panel (a). (c) 1D cuts showing the intensity as function of momentum transfer parallel to the $(00L)$ reciprocal space direction collected at a fixed temperature of 220~K and for several fields ($H\parallel a)$) up to 0.5~T. (d) Field dependence of the wave vector incommensurate modulation obtained from fitting position of the incommensurate peak seen in panel (c). (e) Representative SANS data obtained with the $c$-axis parallel to the beam. This data was collected in zero-field and for a temperature of 420~K.}
\label{fig:SM_SANS}
\end{figure*}

As described in the main manuscript, we collected SANS data for two scattering geometries, 1) with the real space $a$-axis parallel to the beam, which is primarily sensitive to the $(0KL)$ scattering plane, and 2) with the real space $c$-axis parallel to the beam, which predominantly probes the $(HK0)$ scattering plane. The first configuration enabled the observation of the incommensurate out-of-plane modulated state. Here, we report the temperature (Fig.~\ref{fig:SM_SANS}(a,b)) and field (Fig.~\ref{fig:SM_SANS}(c,d)) dependence of this incommensurate modulation. By fitting the position of the incommensurate peak with a Gaussian function, we determined the average modulation vector. The error bars represent the full width at half maximum (FWHM) of the Gaussian fit, roughly quantifying the distribution of modulation vectors. In zero-field (Fig.~\ref{fig:SM_SANS}(b)), the temperature dependence reveals the modulated state exists only between $\sim$ 180~K and $\sim$ 290~K, which is consistent with our large angle neutron diffraction results. The spatial modulation of the wave vector varies slightly from 0.056 $\angstrom^{-1}$ at 180~K to 0.025 $\angstrom^{-1}$ at 290~K. The field dependence, also discussed in the main text, demonstrates that the incommensurate peak vanishes above $\sim$ 0.2~T. Fig.~\ref{fig:SM_SANS}(d) shows a slight decrease in the spatial periodicity of the incommensurate state below 0.2~T. Finally, Fig.~\ref{fig:SM_SANS}(e) shows representative SANS data collected with the $c$-axis parallel to the beam. This scattering geometry was used to estimate the temperature dependence of the out-of-plane ferromagnetic scattering (black square data in Fig.~5(c) of the main manuscript) where the red box in Fig.~\ref{fig:SM_SANS}(e) highlights the precise integration region. 

